\documentclass[%
 reprint,
superscriptaddress,
 amsmath,amssymb,
 aps,
floatfix,
]{revtex4-2}

\usepackage{graphicx}%
\usepackage{dcolumn}%
\usepackage{bm}%

\usepackage{microtype}
\usepackage{subcaption}
\usepackage{float}
\usepackage{placeins}  %
\usepackage{booktabs}
\usepackage{mathtools}
\usepackage{amsthm}
\usepackage[ruled,linesnumbered]{algorithm2e}
\SetKwInput{KwIn}{Require}
\SetKwInput{KwOut}{Ensure}
\usepackage{tikz}
\usetikzlibrary{shapes.geometric, arrows.meta, positioning, fit, backgrounds}
\usepackage{xcolor}
\usepackage{hyperref}
\hypersetup{
  colorlinks=true,
  linkcolor=blue!50!black,
  citecolor=blue!50!black,
  urlcolor=blue!50!black
}
\usepackage[capitalize,noabbrev]{cleveref}
\makeatletter
\AtBeginDocument{%
  \long\def\@makecaption#1#2{%
    \par
    \vskip\abovecaptionskip
    \begingroup
     \small\rmfamily
     \flushing
     \@make@capt@title{#1}{#2}\par
    \endgroup
    \vskip\belowcaptionskip
  }%
}
\makeatother

\usepackage[disable,textsize=tiny]{todonotes}
\theoremstyle{plain}

\theoremstyle{definition}

\theoremstyle{remark}

\begin{document}

\preprint{APS/123-QED}

\title{Designing Quantum Control Protocols via Vibe-Function Reasoning \\ with Large Language Models}

\author{Yusheng Zhao}
\thanks{These authors contributed equally to this work.}
\affiliation{The Hong Kong University of Science and Technology (Guangzhou), Guangzhou, China}

\author{Han Wang}
\thanks{These authors contributed equally to this work.}
\affiliation{The Hong Kong University of Science and Technology (Guangzhou), Guangzhou, China}

\author{Xin Liu}
\thanks{These authors contributed equally to this work.}
\affiliation{Department of Physics, Tsinghua University, Beijing, China}

\author{Xinjie Song}
\affiliation{School of Physics, Peking University, Beijing, China}

\author{Jixi He}
\affiliation{School of Science, Sun Yat-sen University, Shenzhen, China}

\author{Lingwei Song}
\affiliation{Shenzhen University, Shenzhen, China}

\author{Yuanhe Ji}
\affiliation{Department of Physics, Southern University of Science and Technology, Shenzhen, China}

\author{Ken Deng}
\affiliation{Department of Physics, Tsinghua University, Beijing, China}

\author{Runqing Zhang}
\affiliation{China Mobile (Suzhou) Software Technology Co., Ltd., Suzhou, China}

\author{Jize Han}
\affiliation{China Mobile (Suzhou) Software Technology Co., Ltd., Suzhou, China}

\author{Zhiguo Huang}
\affiliation{China Mobile (Suzhou) Software Technology Co., Ltd., Suzhou, China}

\author{Ling Qian}
\affiliation{China Mobile (Suzhou) Software Technology Co., Ltd., Suzhou, China}

\author{Di Luo}
\email[Correspondence: ]{diluo@tsinghua.edu.cn}
\affiliation{Department of Physics, Tsinghua University, Beijing, China}
\affiliation{C.N. Yang Institute of Advanced Study, Tsinghua University, Beijing, China}

\date{\today}

\begin{abstract}
Designing accurate and interpretable quantum control protocols is important for quantum technologies, but conventional numerical solvers can produce protocols that are difficult to interpret and generalize, while large language models (LLMs) may lack the accuracy needed for precise control. We introduce VibeFn-QCtrl, a training-free framework for quantum control protocol design based on vibe-function reasoning, in which an LLM proposes compact analytic control functions and a numerical optimizer refines their coefficients through iterative feedback. We develop QCtrl-Bench, a structured benchmark spanning single- and multi-qubit systems, closed and open quantum dynamics, noiseless and noisy settings, and both analytic and numerical protocols. Across the benchmark, VibeFn-QCtrl achieves performance competitive with or exceeding state-of-the-art conventional solvers and prompt-only LLM baselines in both noiseless and noisy settings.
The method exhibits favorable inference time and pulse resolution scaling behaviours and produces interpretable analytic protocols. Together, these results suggest that VibeFn-QCtrl offers a promising approach to efficient and interpretable quantum control protocol design.
\end{abstract}

\maketitle

\section{Introduction}
\label{sec:main}

\begin{figure*}[t!]
  \centering
  \includegraphics[width=0.8\textwidth]{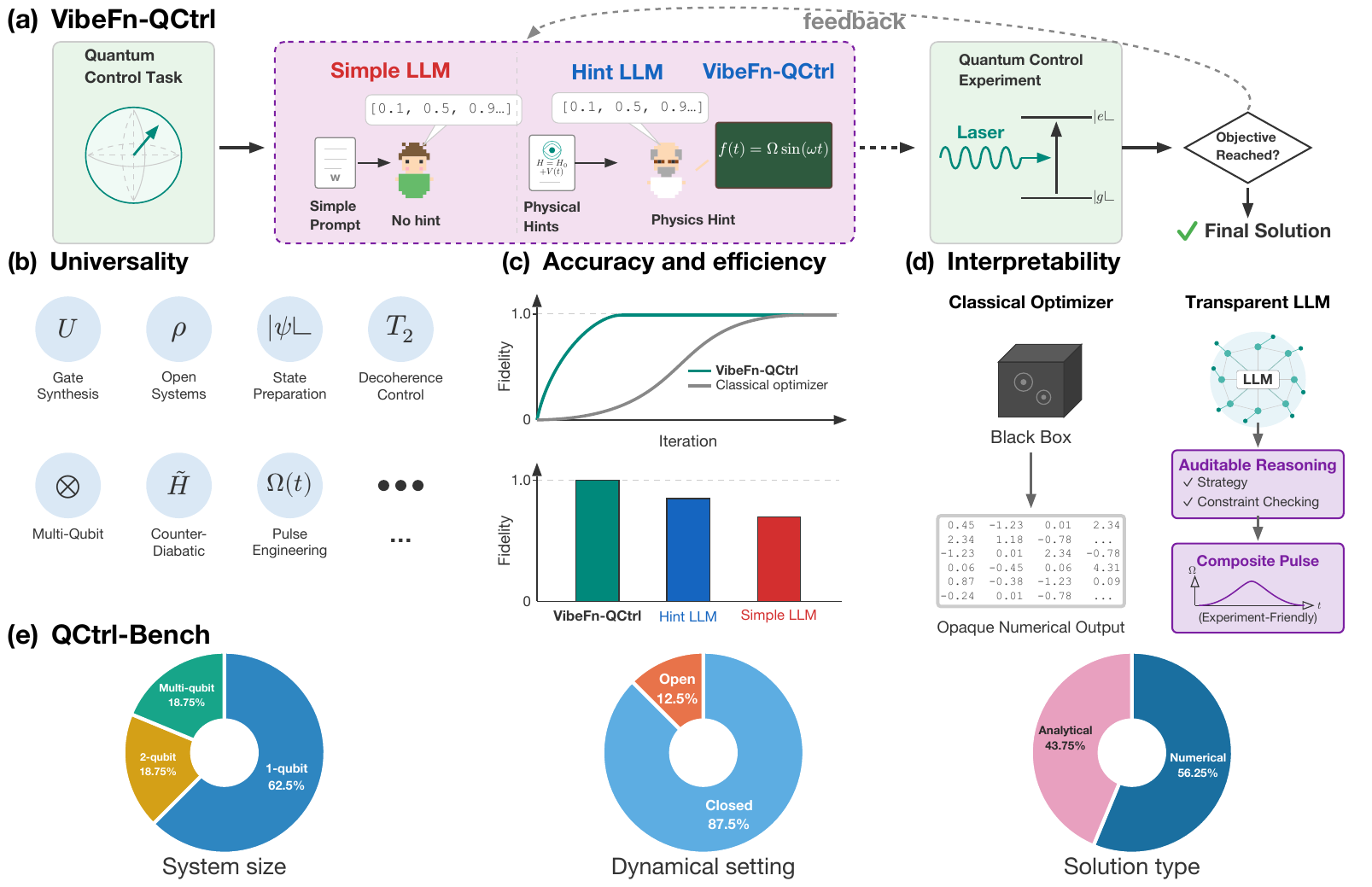}
  \vspace{-4pt}
  \caption{Visual summary of VibeFn-QCtrl and QCtrl-Bench.
    (a)~\emph{VibeFn-QCtrl framework.} VibeFn-QCtrl combines symbolic reasoning with optimization to propose analytic control ans\"atze and refine their parameters through feedback, unlike Simple LLM and Hint LLM which directly generate numerical pulse amplitudes.
    (b)~\emph{Task breadth.} Representative protocol-design tasks across diverse systems and dynamics.
    (c)~\emph{Accuracy and efficiency.} Fidelity and convergence comparisons with classical and prompt-only LLM baselines.
    (d)~\emph{Interpretability.} Analytic control functions make the proposed protocol structure directly inspectable.
    (e)~\emph{QCtrl-Bench composition.} Distribution of benchmark tasks by system size, dynamical setting, and reference-solution type.}
  \label{fig:dataset}
\end{figure*}

Designing quantum control protocols requires choosing time-dependent fields that prepare desired states or implement gates while respecting physical constraints. Such protocols underpin quantum computing, simulation, sensing, and metrology \cite{glaser2015, koch2022}. Advances in superconducting circuits, trapped ions, neutral atoms, and solid-state spins have expanded the scale and controllability of experimental devices \cite{wright2019, bruzewicz2019, krantz2019, browaeys2020, evered2023}, but useful protocols must remain accurate under finite bandwidth and amplitude limits, crosstalk, leakage, and calibration drift. More control channels, longer durations, and finer time resolution enlarge the design space, while device constraints and model mismatch can complicate optimization. The challenge is to design protocols that combine accuracy and robustness with efficient search and interpretable physical structure \cite{dalessandro2007}.

Analytical control theory can reveal the mechanism of a protocol and yield closed-form solutions in selected settings \cite{dalessandro2007}, but extending those solutions to constrained or dissipative systems is often difficult. Numerical methods such as GRAPE, Krotov, and CRAB can reach high fidelities \cite{khaneja2005, krotov1996, doria2011}; however, their pulse representation is generally chosen before optimization, and the resulting protocol may be difficult to interpret or transfer to a new task. CRAB already uses compact functional bases, so compact parameterization alone does not resolve how to choose an appropriate form. Reinforcement learning \cite{niu2019, porotti2019, bukov2018RLphases, metz2023selfcorrecting, reuer2023realtime, zen2025faultTolerant}, evolutionary strategies \cite{judson1992}, neural-network parameterizations \cite{leung2017, vaidhyanathan2026transformer}, and differentiable programming \cite{schaefer2020,PRXQuantum.2.020332, porotti2023feedbackGRAPE} broaden the available design strategies, but can require task-specific choices or training; recent reviews survey these approaches \cite{duncan2025taming, bukov2026RLreview}. Despite this progress, designing protocols across diverse systems that perform well under noise, require few evaluation queries, and remain interpretable is still difficult.

Large language models (LLMs) offer a way to propose such forms from a physical task description. Their demonstrated abilities in mathematical reasoning \cite{trinh2024}, code synthesis \cite{chen2021codex}, scientific problem solving \cite{romera2024}, quantum-circuit design \cite{yang2025qcircuitbench, arlt2026metadesign}, and autonomous scientific experimentation \cite{arlt2025autonomous} suggest that they can combine physical concepts with symbolic expressions. Yet prompting an LLM to emit every pulse amplitude asks it to solve a continuous, high-dimensional optimization problem token by token; this becomes less reliable as pulse resolution and the number of control channels grow.

\paragraph{VibeFn-QCtrl.} We introduce \emph{VibeFn-QCtrl} (VibeFunction Quantum Control; Figure~\ref{fig:dataset}), a framework for designing quantum control protocols through \emph{vibe-function reasoning}. VibeFn-QCtrl contains a \textit{dual-agent} system with an iterative loop: a \emph{proposal agent} designs compact analytic control functions, and a \emph{reflection agent} turns the strengths and weaknesses of earlier proposals into physics-guided feedback for the next round. Between these agents, proposed expressions are checked against a constrained symbolic grammar and parsed in a sandbox, then a coefficient optimizer based on simultaneous-perturbation stochastic approximation (SPSA) tunes their parameters using fidelity feedback \cite{spall1992}. The loop combines the LLM's ability to reason about physical control structures with the numerical optimizer's precision, replacing long pulse arrays with a search over fewer, physically meaningful parameters. It is designed to produce accurate, query-efficient, and interpretable protocols across diverse quantum systems without task-specific training, while remaining effective under calibration errors and as inference effort and pulse resolution increase. The full algorithm is given in Methods, Section~\ref{sec:fspsa} (Algorithm~\ref{alg:fspsa}).

\paragraph{QCtrl-Bench.} To our knowledge, \emph{QCtrl-Bench} is the first structured benchmark for LLM-driven quantum control protocol generation, as distinct from benchmarks of quantum knowledge recall, circuit compilation, or property prediction \cite{lu2022scienceqa, hendrycks2021, li2023qasmbench, quetschlich2023, perrier2022, tang2023}. Its sixteen tasks span single- and multi-qubit systems, closed and open dynamics, and problems with analytic or numerical reference protocols. Each task specifies the physical model, admissible controls, and fidelity objective, with references drawn from GRAPE \cite{khaneja2005}, CRAB \cite{doria2011}, learning-based controllers \cite{leung2017, schaefer2020}, or closed-form solutions. We test protocols both without miscalibration and with fixed $2\%$ offsets in the drift parameters, across four LLMs (Claude Sonnet 4.5, Gemini 3.0 Flash, GPT-OSS-120B, and Qwen 3.5 Plus). This design lets us assess fidelity, query use, pulse-resolution behavior, and the physical interpretability of the resulting protocols; full task specifications and evaluation details appear in Methods, Section~\ref{sec:pipeline}, and Appendix~\ref{app:tasks}. Across QCtrl-Bench, VibeFn-QCtrl matches or exceeds the classical reference on 15 of 16 noiseless tasks and all tasks under fixed calibration offsets. VibeFn-QCtrl shows inference-time scaling through fidelity gains across independent repetitions and pulse-resolution scaling through stable fidelity on finer time grids, while deriving physically interpretable analytic protocols. It also outperforms prompt-only LLM baselines on complex multi-qubit tasks, demonstrating the value of efficient and interpretable LLM-guided protocol design across diverse quantum control problems.

\section{Results}
\label{sec:results}

We compare VibeFn-QCtrl against two classes of baselines: (i) classical numerical controllers, including gradient-based, Fourier-based, deep-learning-based, and gradient-free pulse optimization methods, and (ii) prompt-only LLM baselines that directly generate pulse amplitudes without symbolic form proposal. The prompt-only baselines are divided into two settings: a \emph{Simple LLM} condition that provides only the task specification together with a minimal reasoning instruction, and a more informed \emph{Hint LLM} condition that additionally supplies physics-aware guidance, relevant control paradigms, typical noise mechanisms, and structured feedback from previous iterations. In both settings, the LLM must directly emit one numerical amplitude per time slice and per control channel, without access to coefficient optimization or symbolic protocol refinement. All methods are evaluated under identical environments, fidelity objectives, and noise conditions, allowing QCtrl-Bench to provide a rigorous and controlled framework for studying universality, accuracy, efficiency, robustness, and interpretability in LLM-driven quantum control.

Across QCtrl-Bench, VibeFn-QCtrl demonstrates strong performance in accuracy, robustness, efficiency, scalability, and interpretability. In the noiseless regime ($\sigma = 0$; Section~\ref{sec:noiseless}), VibeFn-QCtrl matches or exceeds the strongest classical baseline on $15$ of $16$ tasks, while consistently outperforming prompt-only LLM baselines. Under calibration noise ($\sigma = 0.02$; Section~\ref{sec:noisy}), where the realized Hamiltonian differs from the nominal system, VibeFn-QCtrl matches or surpasses the corresponding gradient-free classical baseline on all $16$ tasks, including $11$ strict improvements, demonstrating strong robustness against model mismatch. The framework also achieves favorable query efficiency (Section~\ref{sec:efficiency}), requiring substantially fewer evaluations than direct pulse-amplitude optimization. In addition, VibeFn-QCtrl exhibits efficient inference-time scaling and pulse resolution scaling (Section~\ref{sec:scaling}), whereas prompt-only LLM baselines degrade significantly when generating long pulse value vectors. Finally, VibeFn-QCtrl naturally produces interpretable control protocols and frequently derives physically meaningful analytical structures (Section~\ref{sec:interp}), showing that physics-informed symbolic reasoning enables the framework not only to optimize quantum control accurately, but also to uncover compact and human-auditable physical mechanisms rather than merely fitting high-dimensional numerical pulse amplitudes.

\subsection{Accuracy and efficiency}

\subsubsection{Noiseless regime}
\label{sec:noiseless}

\begin{figure}[t]
  \centering
  \includegraphics[width=\columnwidth]{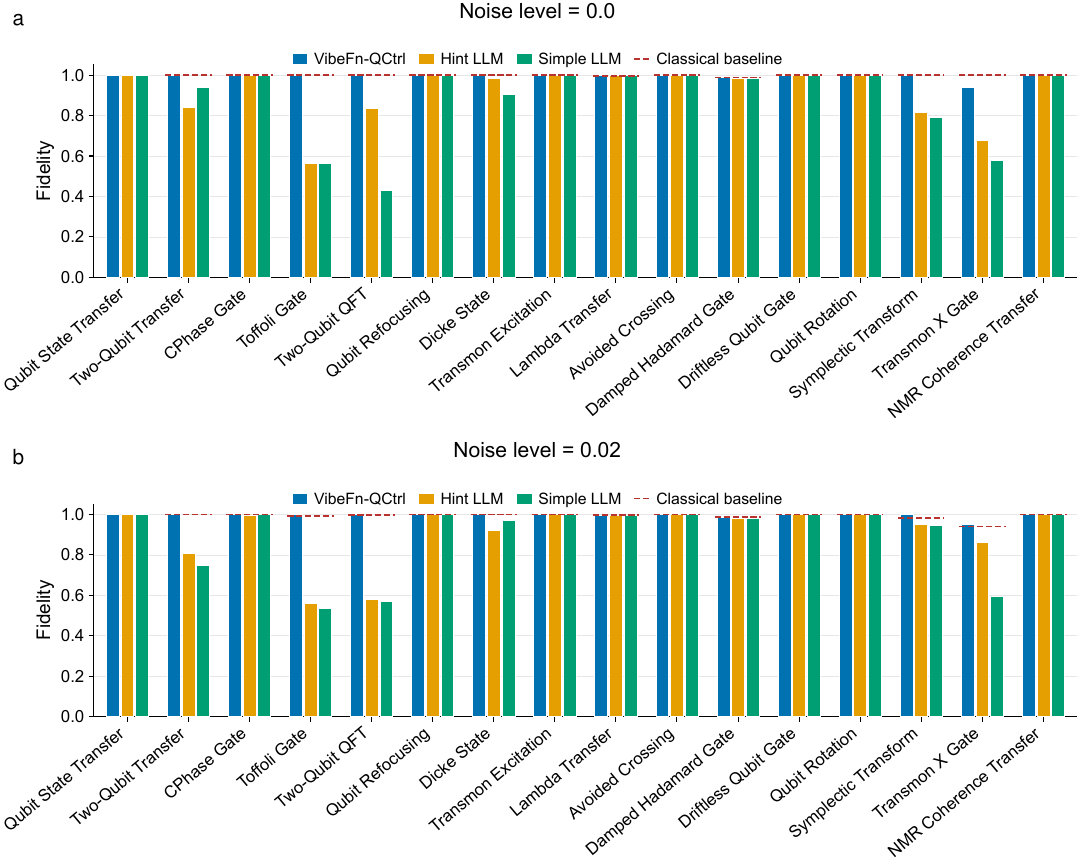}
  \caption{\textbf{Per-task best fidelity on QCtrl-Bench: VibeFn-QCtrl versus classical and pure-LLM baselines.} (a)~Noiseless regime ($\sigma = 0$); the classical baseline (red dashed segment per task) is the established classical optimizer per task (Methods, Section~\ref{sec:baselines}). (b)~Calibration-noise regime ($\sigma = 0.02$); the classical baseline is exposed to the same calibration noise as VibeFn-QCtrl. The details is documented in Methods, Section~\ref{sec:pipeline}.}
  \label{fig:task_characterization}
\end{figure}

Figure~\ref{fig:task_characterization}(a) summarizes the benchmark accuracy in the noiseless setting ($\sigma=0$) across all sixteen QCtrl-Bench tasks. We compare VibeFn-QCtrl against two classes of baselines: established classical baselines for each task reported in the literature (e.g.\ GRAPE, CRAB, Krylov and reinforcement learning; see Methods, Section~\ref{sec:baselines}), and two prompt-only LLM baselines, namely \emph{Simple LLM} and \emph{Hint LLM}.

Across the benchmark, VibeFn-QCtrl consistently achieves high-fidelity control, reaching $F \ge 0.94$ on every task and achieving near-unit fidelity ($F \ge 0.9999$) on nine tasks (Figure~\ref{fig:dataset}b). Compared with classical optimization baselines, VibeFn-QCtrl matches or approaches the strongest classical result on $15$ out of $16$ tasks, with only Transmon Logical X Gate approximately $0.06$ lower relative to GRAPE. Notably, the advantage of VibeFn-QCtrl becomes most apparent on difficult multi-qubit and high-dimensional tasks. For example, on Toffoli Gate Synthesis and Two-Qubit Fourier Gate Synthesis, VibeFn-QCtrl maintains fidelities above $0.999$ by proposing compact physically meaningful ans\"atze, such as layered pulse structures or cosine-envelope protocols, which are subsequently refined by coefficient optimization.

In contrast, the prompt-only LLM baselines degrade substantially as task complexity increases. While both Simple LLM and Hint LLM can solve relatively simple tasks with high fidelity, their performance deteriorates on more challenging settings, including Two-Qubit Fourier Gate Synthesis, Toffoli Gate Synthesis, Coupled-Oscillator Symplectic Transform, and Transmon Logical X Gate, where the fidelity gap can reach nearly $0.6$. Although Hint LLM consistently improves over Simple LLM by incorporating physics-aware guidance and iterative feedback, both baselines remain fundamentally limited by the difficulty of directly generating long numerical pulse vectors. These results demonstrate that the advantage of VibeFn-QCtrl arises not merely from providing physics hints to the model, but from enabling symbolic control-form proposal combined with efficient coefficient refinement. Furthermore, VibeFn-QCtrl provides interpretable formula for quantum control protocol, which will be presented in Section~\ref{sec:interp}.

\subsubsection{Noisy regime}  
\label{sec:noisy}

To evaluate robustness under realistic hardware imperfections, we study all QCtrl-Bench tasks in a calibration-noise setting with $\sigma=0.02$, where the drift Hamiltonian is perturbed by fixed multiplicative offsets applied independently to task-specific physical parameters such as Rabi frequencies, couplings, and energy gaps:
\begin{equation}
p_{\mathrm{noisy}} = p (1 + \delta_p), \qquad \delta_p \in \{-\sigma, +\sigma\}, \qquad \sigma=0.02.
\label{eq:calibration_noise}
\end{equation}
The perturbation remains fixed throughout each optimization repetition, corresponding to static miscalibration rather than stochastic shot noise. Importantly, each method only knows the nominal Hamiltonian and has no direct access to the perturbed parameters or the value of $\sigma$. VibeFn-QCtrl adapts to the noisy environment indirectly through iterative feedback. At each round, the proposal agent generates a compact analytic protocol, while the resulting fidelity under the perturbed Hamiltonian is returned through the rolling feedback memory together with the rendered control form. The reflection agent then summarizes useful physical patterns and failure modes from previous repetitions, allowing subsequent proposals to progressively adapt to the effective noisy dynamics despite never explicitly observing the calibration offsets.

Figure~\ref{fig:task_characterization}(b) reports per-task best fidelity at the calibration noise $\sigma = 0.02$: Across all sixteen tasks, VibeFn-QCtrl matches or exceeds the classical baseline , achieving $11$ strict improvements while remaining within $0.001$ fidelity on the remaining tasks, and exceeding the two prompt-only LLM baselines on every multi-qubit and high-dimensional task. The advantage comes from searching a low-dimensional symbolic ansatz rather than a flat $\mathcal{O}(N_t\cdot C)$-dimensional amplitude vector ($N_t$ piecewise-constant time slices, $C$ independent control channels), so VibeFn-QCtrl is more drift-robust than the classical baseline and more reliable than naive amplitude generation.
The largest gains over the classical baseline are on Dissipative Lambda Transfer ($F = 0.997$ vs.\ $0.966$, $\Delta F = +0.031$), Coupled-Oscillator Symplectic Transform ($0.9997$ vs.\ $0.983$, $+0.017$), Transmon Logical X Gate ($0.950$ vs.\ $0.941$, $+0.010$), and Toffoli Gate Synthesis ($0.9996$ vs.\ $0.991$, $+0.008$); Transmon Logical X Gate is particularly informative because in the noiseless regime it was the worst VibeFn-QCtrl task, trailing GRAPE by $0.06$ in fidelity, yet under miscalibration the symbolic ansatz now wins, since a few-coefficient analytic form absorbs drift that a full amplitude search cannot.
VibeFn-QCtrl's fidelity advantage over the prompt-only baselines is largest on multi-qubit and high-dimensional tasks: on Two-Qubit Fourier Gate Synthesis and Toffoli Gate Synthesis, Simple LLM and Hint LLM cluster at $F \approx 0.53$--$0.58$ while VibeFn-QCtrl stays above $F = 0.999$; on Two-Qubit State Transfer they reach only $F \approx  0.74$--$0.80$ versus VibeFn-QCtrl's $F = 1.000$; on shorter single-qubit tasks and Avoided-Crossing State Transfer they remain competitive, because direct numerical pulse generation is stable when the pulse sequence is short.

\begin{figure}[t!]
  \centering
  \includegraphics[width=\columnwidth]{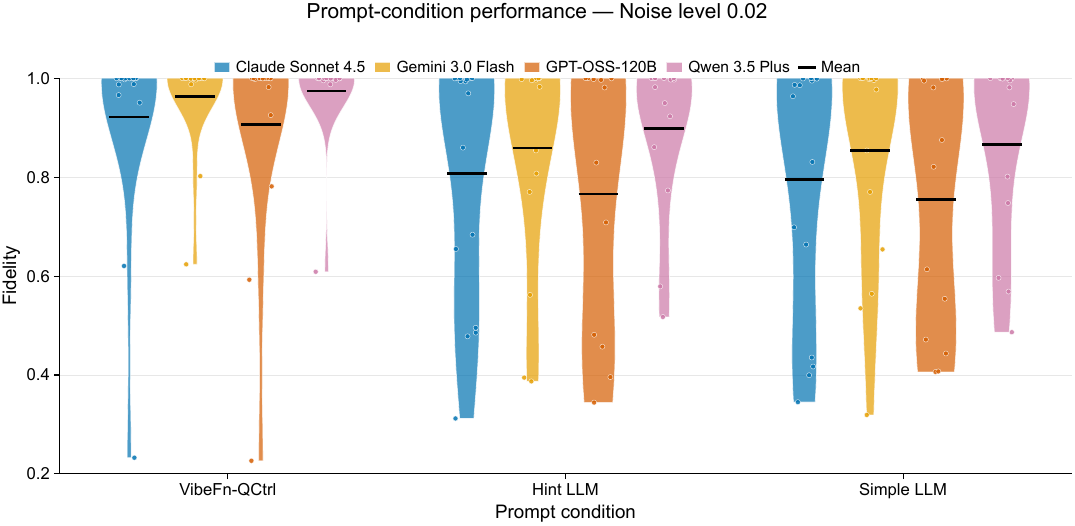}
  \caption{\textbf{Distribution of per-task best fidelity at $\sigma = 0.02$ across LLM-driven methods.} Each violin shows the 16 per-task best fidelities for one (model, method) cell: four models (Claude Sonnet 4.5, Gemini 3.0 Flash, GPT-OSS-120B, Qwen 3.5 Plus) $\times$ three methods (VibeFn-QCtrl, Hint LLM, Simple LLM), giving 12 violins. Per-task fidelity convention: for VibeFn-QCtrl, the best-of-iteration value from a single repetition per task ($n = 1$); for Hint LLM and Simple LLM, the best across $n = 5$ independent repetitions per task. Overlaid scatter points show the 16 individual per-task fidelities; the horizontal black tick inside each violin marks the mean. VibeFn-QCtrl violins concentrate near $F = 1$ across all four models; Hint LLM and Simple LLM violins are broader and run longer low-fidelity tails.
  The $\sigma = 0$ case is shown in Figure~\ref{fig:prompt_fidelity_combined}.}
  \label{fig:prompt_barplot_noise002}
\end{figure}

Figure~\ref{fig:prompt_barplot_noise002} stratifies the same $\sigma = 0.02$ results by the underlying LLM, showing per-task fidelity distributions for each (method, model) cell. VibeFn-QCtrl violins concentrate near $F = 1$ with per-model mean fidelity in the $0.91$--$0.98$ range across all four models, while Hint LLM and Simple LLM violins broaden into long low-fidelity tails reaching $F \approx 0.3$--$0.5$ with per-model means dropping to $0.75$--$0.90$; symbolic protocol proposal thus compresses the cross-model spread from approximately $0.12$ in fidelity under the prompt-only baselines to $0.06$ under VibeFn-QCtrl. Across all three methods the four LLMs rank consistently as Qwen 3.5 Plus $>$ Gemini 3.0 Flash $>$ Claude Sonnet 4.5 $>$ GPT-OSS-120B, but VibeFn-QCtrl absorbs most of this cross-model capability gap: the weakest model under VibeFn-QCtrl (GPT-OSS-120B, mean $F = 0.91$) matches the strongest model under Hint LLM (Qwen 3.5 Plus, mean $F = 0.90$). The low-fidelity tails of the prompt-only methods are heaviest on Claude Sonnet 4.5 and GPT-OSS-120B and are driven by the same multi-qubit tasks (Two-Qubit Fourier Gate Synthesis, Toffoli Gate Synthesis) that fail in Figure~\ref{fig:task_characterization}(b).

Beyond final fidelity, VibeFn-QCtrl is more efficient than the baselines in both iterations and language-model tokens. Under calibration noise ($\sigma = 0.02$), VibeFn-QCtrl reaches $F = 0.999$ in fewer iterations than the classical baseline on tasks that admit a compact analytical form(Figures~\ref{fig:efficiency}), and consumes fewer or comparable LLM tokens than both prompt-only baselines (Hint LLM and Simple LLM) across all four evaluated models(Figure~\ref{fig:prompt_tokens_combined}). Per-task and per-model breakdowns are given in Appendix~\ref{sec:efficiency}.

\subsection{Inference-time and pulse-resolution scaling}
\label{sec:scaling}

Inference-time scaling and pulse-resolution scaling provide two complementary perspectives on the behavior of VibeFn-QCtrl. The first examines how performance improves as the framework is allowed more independent proposal-reflection repetitions, probing whether additional reasoning cycles can continue to refine control quality. The second studies how performance changes as the number of piecewise-constant time slices increases, which directly tests whether the framework remains stable in high-dimensional control settings. Together, these scaling tasks highlight a key distinction between VibeFn-QCtrl and prompt-only LLM approaches: VibeFn-QCtrl operates through compact symbolic control representations whose complexity remains largely independent of pulse discretization, whereas naive LLM baselines must directly generate increasingly long numerical pulse vectors as the control resolution grows.

\begin{figure}[t!]
  \centering
  \includegraphics[width=\columnwidth]{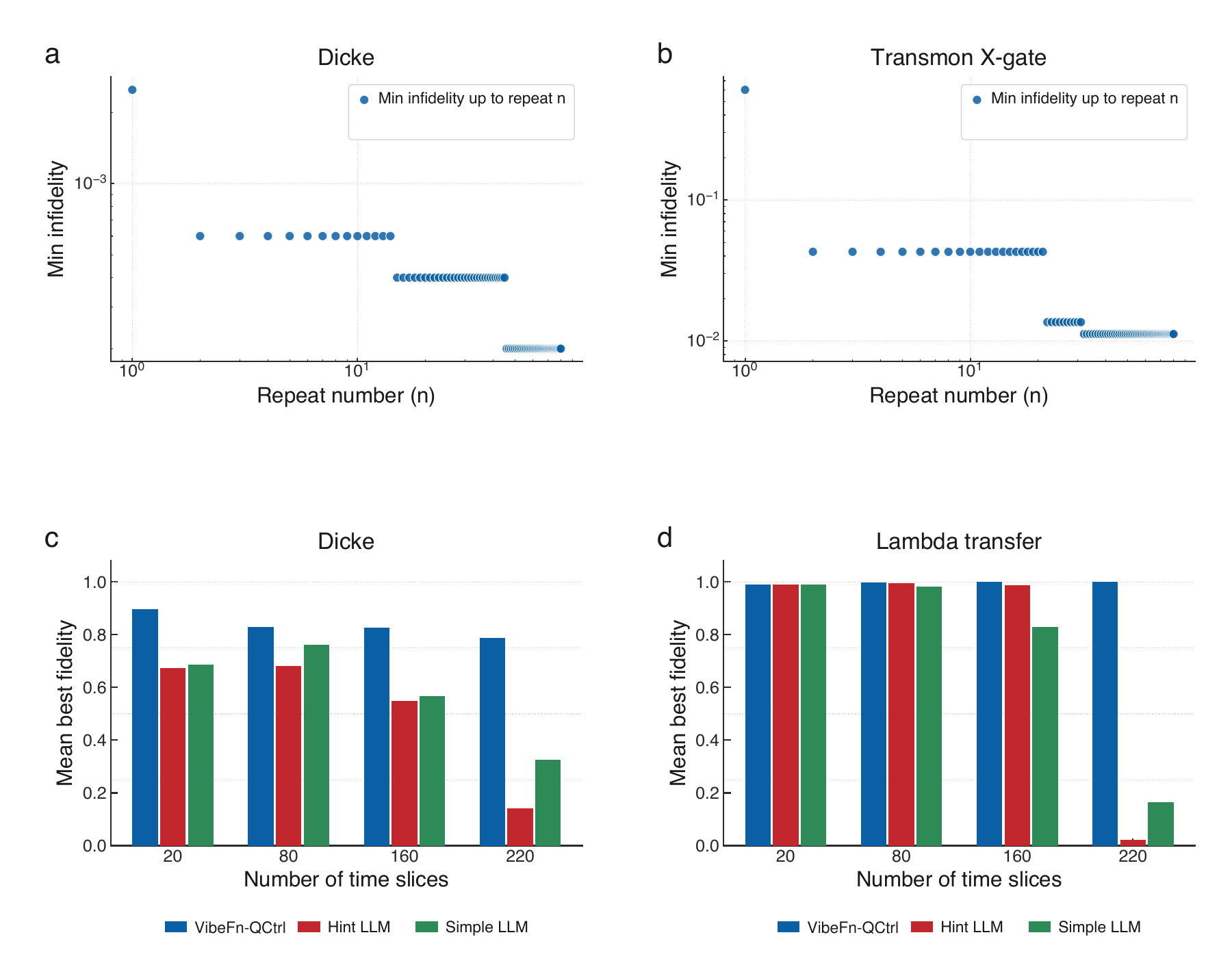}
  \caption{\textbf{Scaling diagnostics.}
  (a,\,b)~VibeFn-QCtrl running-minimum infidelity $\epsilon_n = \min_{i \le n}(1 - F_i)$ versus the number of independent repetitions $n$. On both tasks the running-minimum infidelity decreases with $n$, reaching $\sim\!10^{-4}$ for Dicke State Preparation and $\sim\!10^{-2}$ for Transmon Logical X Gate.
  (c,\,d)~Mean best fidelity plotted against the number of piecewise-constant time slices. For both Dicke State Preparation and Dissipative Lambda Transfer tasks, VibeFn-QCtrl maintains high fidelity while pure LLM methods deteriorate rapidly with increasing $N_t$. All tasks in the figures are performed under calibration noise $\sigma = 0.02$.}
  \label{fig:scaling_law_combined}
\end{figure}

Figure~\ref{fig:scaling_law_combined}(a--b) shows the \textit{inference-time scaling} behavior of VibeFn-QCtrl under calibration noise ($\sigma=0.02$) on representative tasks including Dicke State Preparation and Transmon Logical X Gate. On both tasks the running-best infidelity decreases in discrete stepwise drops as the number of independent repetitions $n$ increases, reflecting the progressive discovery of better control ans\"atze across repetitions. Dicke State Preparation descends to infidelities on the order of $10^{-4}$, while Transmon Logical X Gate plateaus near $10^{-2}$; in both cases, additional repetitions continue to deliver fidelity improvements rather than stalling at the first solution. Overall, the observed trend demonstrates that VibeFn-QCtrl benefits naturally from increased inference-time compute through iterative reasoning and protocol refinement rather than relying solely on larger pretrained model capacity.

Figure~\ref{fig:scaling_law_combined}(c--d) presents the \textit{pulse-resolution scaling} results as the number of time slices increases from $N_t=20$ to $N_t=220$. VibeFn-QCtrl maintains consistently strong fidelities across all tested resolutions, with only modest variation as the discretization grid becomes finer. In contrast, both Simple LLM and Hint LLM baselines deteriorate rapidly with increasing $N_t$, particularly at high resolutions where the models must generate extremely long numerical pulse sequences directly. At $N_t=220$, the fidelities of the prompt-only baselines fall below $0.4$ on representative tasks, whereas VibeFn-QCtrl remains stable. This difference arises because VibeFn-QCtrl searches over compact analytic control forms whose effective complexity remains nearly independent of the pulse grid, while prompt-only methods scale directly with the length of the numerical control vector. These results identify pulse-resolution scaling as a central advantage of symbolic protocol proposal and demonstrate that VibeFn-QCtrl can remain effective in regimes where direct high resolution pulse generation becomes unstable.

\subsection{Interpretability}
\label{sec:interp}

\begin{figure}[t]
  \centering
  \includegraphics[width=\columnwidth]{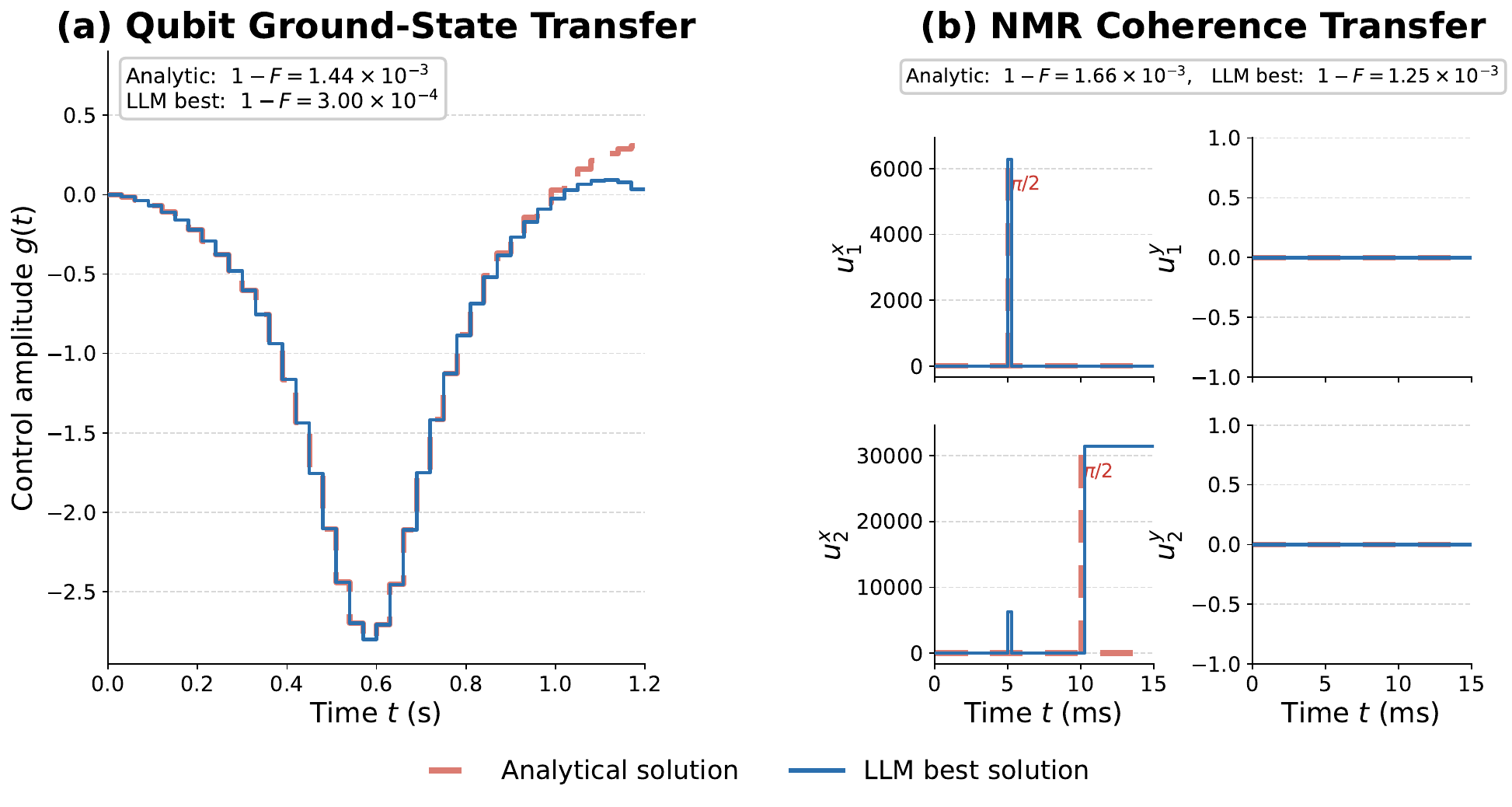}
  \caption{Analytical structure in VibeFn-QCtrl outputs at noiseless regime. (a) Qubit Ground-State Transfer: the LLM derives the continuous-time CD control law and identifies discretization-dependent corrections. (b) NMR Coherence Transfer: the LLM utilizes the product-operator evolution to derive four control pulses with high fidelity.}
  \label{fig:fitting_tasks}
\end{figure}

Beyond final performance, VibeFn-QCtrl offers interpretability at two distinct levels:
(i) \emph{Formula-Level Interpretability}, as it returns explicit, compact functional expressions instead of opaque numerical pulse arrays; and
(ii) \emph{Reasoning-Level Interpretability}, since it exposes the LLM's step-by-step reasoning trace explaining how those expressions were chosen and physically derived.
We can therefore inspect not only whether the proposed controls correspond to valid physical mechanisms, but also whether the model's intermediate reasoning reflects the correct underlying physics.
To understand the LLM's contribution, we analyze two tasks from VibeFn-QCtrl with zero-step coefficient-optimizer, where the LLM proposes both the functional form and its initial coefficients directly, with no subsequent numerical refinement. We note that these tasks are curated such that their exact setup has not appeared in the literature, and the closed-form references are never exposed to the model; the prompt provides only the Hamiltonian, available controls, objective, and time discretization.
As illustrated in Fig.~\ref{fig:fitting_tasks}, even without parameter optimization, VibeFn-QCtrl protocols achieve high physical fidelity by deriving the correct mathematical structures.

The first case is Qubit Ground-State Transfer, evaluated on Qwen 3.5 Plus. In this task, the goal is to steer a spin-1/2 particle along its instantaneous ground state under a time-dependent qubit Hamiltonian:
\begin{equation}
\begin{aligned}
    H_{\mathrm{total}}(t) &= 1.0\!\times\!\left(1 + (t/1.2)^3\right)\sigma_x \\
    &\quad + \left(2.0 - 4.0 \times S(t/1.2)\right)\sigma_z - g(t)\sigma_y
\end{aligned}
    \label{eq:qubit_hamiltonian}
\end{equation}
where $S(\tau) = 10\tau^3 - 15\tau^4 + 6\tau^5$ is a quintic smoothstep function, and the cubic transverse (i.e., $1.0\!\times\!\left(1 + (t/1.2)^3\right)$) and smoothstep longitudinal (i.e., $2.0 - 4.0 \times S(t/1.2)$) sweeps form a custom configuration designed specifically for this benchmark that has not appeared in prior literature. The control $g(t)$ is discretized into 40 slices over $T=1.2$.

\noindent\emph{LLM-derived formula.}
The LLM actively recognizes the control channel coefficients in $H_{\mathrm{total}}(t)$ to derive the continuous-time counter-diabatic field, introducing $\Omega_x(t)$ and $\Omega_z(t)$ as the coefficients of the Pauli operators $\sigma_x$ and $\sigma_z$:
\[
g(t)=\frac{\Omega_x\dot{\Omega}_z-\Omega_z\dot{\Omega}_x}
{2(\Omega_x^2+\Omega_z^2)}.
\]
As shown in Fig.~\ref{fig:fitting_tasks}(a), this derived formula is closely aligned with the analytical counter-diabatic driving protocol. Indeed, the LLM further introduced small correction terms on top of the counter-diabatic term in an attempt to compensate for errors introduced by temporal discretization, achieving higher fidelity state preparation than the normal CD driving protocol without any parameter optimization.

\noindent\emph{LLM reasoning.} 
The LLM's reasoning trace (see Appendix~\ref{app:trace_cd} for the verbatim trace) demonstrates clear physical intuition by mapping the drifting qubit Hamiltonian (the uncontrolled terms along $\sigma_x$ and $\sigma_z$ in Eq.~\eqref{eq:qubit_hamiltonian}: $H_0(t) = 1.0\!\times\!\left(1 + (t/1.2)^3\right)\sigma_x + \left(2.0 - 4.0 \times S(t/1.2)\right)\sigma_z$) to a spin-1/2 particle in an effective magnetic field $\vec{B}_0(t) = (\Omega_x(t), 0, \Omega_z(t))$. For instance, it derives: ``The CD term for a spin in a magnetic field $B_0(t)$ is $H_{\mathrm{CD}} = (B_0 \times \dot{B}_0) / (2 |B_0|^2)$''. Furthermore, it actively diagnoses and identifies the temporal discretization effects introduced by the finite time grid ($N_t=40$), pointing out in its reasoning trace that the ``Midpoint rule is 2nd order'' (implying the global simulation error of the time-ordered propagator scales as $\mathcal{O}(\Delta t^2)$) and explaining the physical reason why the discrete sequence of rotations fails to perfectly track the continuous-time analytical CD drive (e.g., ``In discrete time, the 'kick' from $H(t_k)$ is exact for that slice, but the sequence of rotations doesn't perfectly track'').

NMR Coherence Transfer provides a sequence-level example, evaluated on Qwen 3.5 Plus. This task considers polarization transfer between two weakly coupled spin-1/2 particles (labeled 1 and 2) under a J-coupling Hamiltonian:
\begin{equation}
    H(t) = 2\pi J S_1^z S_2^z + u_1^x(t)S_1^x + u_1^y(t)S_1^y + u_2^x(t)S_2^x + u_2^y(t)S_2^y
\end{equation}
where $J = 100\text{ Hz}$, over a total duration $T = 15.0\text{ ms}$ discretized into exactly 60 time slices ($\Delta t = 0.25\text{ ms}$).

\noindent\emph{LLM-derived formula.}
The LLM reasons that the coupling term $2\pi J S_1^z S_2^z$ generates anti-phase coherence, and utilizes the analytic product-operator evolution under this term for duration $\tau$:
\begin{equation}
    S_2^x \xrightarrow{2\pi J S_1^z S_2^z \tau} S_2^x\cos(\pi J\tau)+2S_1^z S_2^y\sin(\pi J\tau).
    \label{eq:coherence_evolution}
\end{equation}
To maximize the transfer from the single-spin coherence $S_2^x$ to the anti-phase coherence $2 S_1^z S_2^y$ (the second term in Eq.~\eqref{eq:coherence_evolution}), the model targets a state where $\sin(\pi J\tau) = 1$ and $\cos(\pi J\tau) = 0$. Based on this relation, the model analytically derives the optimal delay time $\tau = 1/(2J) = 5\text{ ms}$ (for $J=100\text{ Hz}$), yielding a precise sequence-level control strategy. To translate this into continuous control fields, the LLM proposes Heaviside-based rectangular pulses along the $x$-direction channels $u_1^x(t)$ and $u_2^x(t)$ (in $\text{rad/s}$) while setting $u_1^y(t) = u_2^y(t) = 0$ since no $y$-axis rotations are required for this sequence. Specifically, after the initial $5\text{ ms}$ delay, it applies simultaneous $\pi/2$ pulses of opposite signs (amplitude $-2\pi \times 10^3\text{ rad/s}$ on particle 1 to rotate $S_1^z \to -S_1^y$, and $2\pi \times 10^3\text{ rad/s}$ on particle 2 to rotate $S_2^y \to S_2^z$) over $0.25\text{ ms}$, yielding the exact pulse area of $\pi/2\text{ rad}$ for phase-coherent transfer $2S_1^z S_2^y \to -2S_1^y S_2^z$. Following a second $5\text{ ms}$ delay to obtain target $S_1^x$ at $t=10.25\text{ ms}$, the model addresses the remaining simulation time by applying a strong continuous-wave decoupling field ($2\pi \times 5 \times 10^3\text{ rad/s} \gg 2\pi J$) on $u_2^x(t)$ from $10.25$ to $15.0\text{ ms}$. This spin-locks the second particle ($\langle S_2^z \rangle \approx 0$), effectively suppressing the coupling interaction and freezing the target state at high fidelity, as shown in Fig.~\ref{fig:fitting_tasks}(b).

\noindent\emph{LLM reasoning.}
The LLM's reasoning trace (see Appendix~\ref{app:trace_twospin} for the verbatim trace) illustrates how it maps the continuous-time physical process to the discrete simulation grid while acknowledging the limits of physical approximations. The LLM identifies the problem as ``coherence transfer via scalar coupling'' and constructs the ideal switching sequence (free evolution, phase-appropriate local $\pi/2$ rotations, and a second free-evolution interval). It then translates this to the temporal discretization, mapping the analytical delay $\tau = 5\text{ ms}$ to exactly 20 time slices of $\Delta t = 0.25\text{ ms}$, demonstrating its capacity to perform temporal discretization mapping. It further recognizes that finite-width pulses are imperfect from the delta pulse because ``During the pulse, J-coupling is also active'', introducing side effects from the other term. It also explicitly reasons about how to freeze the target state during the remaining $4.75\text{ ms}$ of the simulation, writing: ``We need to preserve Ix from 10.25 to 15.0. Apply CW decoupling on S (u\_Sx = large). This averages IzSz to 0 (in the frame of S). Actually, if we lock S along x, = 0. So IzSz averages to 0. Ix is stable.'' (here the LLM's symbols $I$ and $S$ map to particle 1 and particle 2 in our notation, so that $I_x$, $u_{Sx}$, and $I_z S_z$ correspond to $S_1^x$, $u_2^x$, and $S_1^z S_2^z$, respectively).

Together, these examples show that interpretability lies in the derivational structure of the solution, not merely in the final pulse representation. In Qubit Ground-State Transfer, the LLM derives a continuous-time CD control law. In NMR Coherence Transfer, it derives an analytical sequence-level mechanism. This hierarchy of formula-level and reasoning-level interpretability is precisely what is lost in raw GRAPE or CRAB amplitude arrays: the numerical pulse may be accurate, but the mechanism is not explicitly named, derived, or checked against the physics.

\section{Discussion}
\label{sec:conclusion}

Designing quantum control protocols across different systems requires both a useful control structure and accurate parameter tuning. We introduced \emph{VibeFn-QCtrl} (VibeFunction Quantum Control), which uses vibe-function reasoning with a large language model to propose compact, physically motivated analytic control functions and a gradient-free optimizer to refine their coefficients using fidelity feedback. Its proposal and reflection agents revise the functional structure in response to earlier results, combining physical reasoning with numerical precision. We evaluated this design strategy on \emph{QCtrl-Bench}, sixteen tasks spanning single- and multi-qubit systems, closed and open dynamics, and both noiseless and fixed calibration-offset conditions. VibeFn-QCtrl matches or exceeds the task-specific classical reference on fifteen of sixteen noiseless tasks and all sixteen tasks with calibration offsets, while outperforming prompt-only LLM baselines by up to nearly $0.6$ in fidelity on selected multi-qubit tasks. It also shows favorable query efficiency, gains from additional independent attempts, and stable fidelity as pulse resolution increases. Its analytic protocols make the proposed physical control mechanisms easier to inspect. Together, these results suggest VibeFn-QCtrl as an efficient and interpretable approach to quantum control protocol design without task-specific training.

Since VibeFn-QCtrl is able to operate through iterative feedback rather than requiring differentiable models, the framework naturally interfaces with hardware-in-the-loop optimization settings. An important next step will be to consider deployment on real quantum hardware.  This opens the possibility of directly applying VibeFn-QCtrl to superconducting quantum processors, neutral-atom arrays, trapped-ion systems, and other programmable quantum platforms subject to calibration drift, analog bandwidth constraints, crosstalk, and device-specific noise. More broadly, VibeFn-QCtrl suggests a path toward integrating scientific reasoning, symbolic structure design, and numerical optimization within a unified agentic framework for scientific machine learning. Extending these ideas to larger many-body quantum systems, hardware-aware optimal control, and autonomous experiment design represents an important opportunity for future research.

\section{Methods}
\label{sec:methods}
\label{sec:dataset}

This section defines VibeFn-QCtrl as a control-synthesis method (Section~\ref{sec:fspsa}), describes the QCtrl-Bench task suite used to evaluate it (Section~\ref{sec:bench_suite}), the classical and pure-LLM baselines we compare against (Section~\ref{sec:baselines}), and the evaluation pipeline and metrics (Section~\ref{sec:pipeline}). The full task descriptions are given in Appendix~\ref{app:tasks}.

\subsection{VibeFn-QCtrl}
\label{sec:fspsa}

VibeFn-QCtrl is an LLM-based quantum control framework built on vibe-function reasoning, designed around the complementary strengths and weaknesses of large language models. While LLMs possess strong high-level physical reasoning and pattern abstraction abilities, they are unreliable for direct high-dimensional numerical optimization. VibeFn-QCtrl therefore assigns the LLM a structural role: instead of generating discretized pulse amplitudes directly, the model is asked to propose compact analytic vibe functions for each control channel, consisting of symbolic functions of time with a small number of freely varying coefficients. The pipeline is organized as a dual-agent iterative optimization loop with three components acting in sequence at each round. A \emph{proposal agent} generates, for every control channel, a parameterized analytic ansatz together with initial coefficient values. A \emph{coefficient optimizer} then numerically refines those coefficients against a fidelity oracle using a gradient-free black-box algorithm; throughout this work we use Spall's simultaneous-perturbation stochastic approximation (SPSA)~\cite{spall1992} for its sample-efficiency, although the framework is compatible with other gradient-free optimizers. Finally, a \emph{reflection agent} distills the round's reasoning trace and observed fidelity into a compact summary that is fed back into the next round's proposal agent. The loop terminates once the optimized infidelity falls below a target tolerance $\varepsilon$ or the outer-loop budget $K$ is exhausted.

The proposal agent reads a structured prompt with three components: (i) a task specification containing the Hamiltonian, control channels, optimization target, physical constraints, and fidelity tolerance; (ii) a physics-guidance block summarizing relevant control paradigms and reminding the model of potential calibration noise or Hamiltonian mismatch; and (iii) a rolling feedback memory, written by the reflection agent, containing selected previous repetitions together with their analytic forms, optimized infidelities, and summaries of why particular strategies succeeded or failed. The first iteration is initialized from a zero-amplitude warm start with an empty feedback memory and asks the model to propose control structures from scratch. At each iteration $k$, the agent outputs, for each control channel, a structured representation
\[
  \big\{ \mathtt{expression}\!: s_c,\; \mathtt{parameters}\!: \{a_1\!:\!\theta_1^{(0)}, \dots, a_p\!:\!\theta_p^{(0)}\}\big\},
\]
where $s_c$ is a symbolic expression in time $t$, total duration $T$, and trainable parameters $\{a_i\}$, and $\{\theta_i^{(0)}\}$ are the agent's proposed initial values. To ensure robustness and physical interpretability, the symbolic grammar is intentionally constrained: only the operations $+,\,-,\,\times,\,/,\,\wedge$ and the functions $\sin,\cos,\exp,\log,\mathrm{erf},\tanh,\mathrm{sinc},\Theta$ are permitted, with at most $p=60$ freely varying coefficients and being non-zero. For example, on a Landau--Zener task the agent may propose a monotonic sweep supplemented by a corrective oscillatory term, whereas for a DRAG-style protocol it may generate a Gaussian envelope together with derivative compensation. Crucially, the agent proposes only the functional structure and reasonable initial coefficient values; all numerical refinement is handled by the coefficient optimizer. This constrained symbolic representation lets VibeFn-QCtrl search efficiently over physically meaningful protocol families rather than unrestricted numerical pulse spaces.

The coefficient optimizer uses SPSA because it requires only two objective evaluations per optimization step regardless of the parameter dimension $p$. The update rule is
\begin{equation}
\begin{aligned}
  &\Delta_k \sim \mathrm{Rademacher}(\{-1,+1\})^p, \qquad y_\pm = J\!\left(\theta_k \pm c_k\,\Delta_k\right), \\
  &\hat g_k = \frac{y_+ - y_-}{2 c_k}\,\Delta_k, \qquad
   \theta_{k+1} = \mathrm{clip}\!\left(\theta_k - a_k\,\hat g_k\right),
\end{aligned}
\label{eq:spsa}
\end{equation}
with gain schedules $a_k = a/(k+1+A)^{0.602}$ and $c_k = c/(k+1)^{0.101}$ following the standard SPSA prescription ($A = 50$; $a$ and $c$ are task-dependent hyperparameters). The objective is $J(\theta)=1-F(\theta)$: at each evaluation, the proposed analytic expression is discretized on the midpoint time grid, converted into a piecewise-constant control sequence, and evaluated by a \emph{fidelity oracle}. In this work the fidelity oracle is a quantum simulator, but it is treated as a black-box evaluator, so the same optimizer applies uniformly across closed-system unitary evolution, Lindbladian open-system dynamics, noisy calibration models, and, in principle, a fidelity estimator running on real hardware. The optimizer runs for at most $B_{\mathrm{opt}}$ steps per round (default $10^3$) and returns the refined infidelity $1-F_k$ together with the rendered analytic protocol obtained by substituting the optimized coefficients into the symbolic expression.

Between rounds, the reflection agent analyzes the previous proposal and its observed fidelity improvement $F_k - F_{k-1}$ and emits a compact set of feedback bullets that preserve useful physical reasoning while avoiding prohibitively long reasoning traces. These bullets, together with that round's analytic form, optimized infidelity, and rendered protocol, are appended to the rolling feedback memory and consumed by the proposal agent in the next iteration. The proposal agent may then either refine the same functional basis or mutate the structure itself by introducing additional oscillatory terms, phase shifts, Gaussian corrections, or alternative basis functions, evolving protocol structures based on distilled physical insight rather than raw numerical amplitudes. This mechanism lets the LLM focus on proposing meaningful structural changes(oscillatory corrections for adiabatic sweeps, derivative compensation for leakage suppression, counter-intuitive pulse ordering in STIRAP-like dynamics), while numerical refinement is handled separately by the coefficient optimizer. The outer loop runs for at most $K$ iterations (default $50$) or until the target tolerance $\varepsilon$ is reached; the complete VibeFn-QCtrl pipeline is summarized in Algorithm~\ref{alg:fspsa}.

\begin{algorithm}[t]
\caption{VibeFn-QCtrl control synthesis.}
\label{alg:fspsa}
\KwIn{Task $\mathcal T$; physics hint $h$; proposal agent $\pi$; reflection agent $\pi'$; outer budget $K$; coefficient-optimization budget $B_{\mathrm{opt}}$; tolerance $\varepsilon$.}
\KwOut{Best ansatz $(s^\star,\,\theta^\star)$ and infidelity $1-F^\star$.}
$\mathcal H \gets \emptyset$;\ \ $F^\star \gets 0$ \tcp*[r]{history and best fidelity}
\For{$k = 1,\,2,\,\ldots,\,K$}{
  $P_k \gets \textsc{BuildPrompt}(\mathcal T,\,h,\,\mathcal H)$\;
  $\{s_c,\,\theta_c^{(0)}\}_c \gets \pi(P_k)$ \tcp*[r]{proposal agent: LLM proposes form + initial coefficients}
  validate $s \gets \{s_c\}_c$ against the schema\;
  $(\theta_k^\star,\,F_k) \gets \textsc{Optimize}\bigl(\,\theta \mapsto 1 - F(\mathcal T,\,s(\theta)),\ \theta^{(0)},\ B_{\mathrm{opt}}\,\bigr)$ \tcp*[r]{Eq.~\eqref{eq:spsa}}
  append $(s,\,\theta_k^\star,\,F_k,\,\text{rendered form})$ to $\mathcal H$; update $F^\star$ and $(s^\star,\theta^\star)$\;
  \If{$k \ge 2$}{
    $r_k \gets \pi'\bigl(\text{reasoning}_{k-1},\ F_k - F_{k-1}\bigr)$; append $r_k$ to $\mathcal H$ \tcp*[r]{reflection agent summarizes useful physical insights}
  }
  \If{$1-F_k \le \varepsilon$}{
    \textbf{break} \tcp*[r]{tolerance reached}
  }
}
\Return $(s^\star,\,\theta^\star,\,1 - F^\star)$\;
\end{algorithm}

\subsection{QCtrl-Bench}
\label{sec:bench_suite}

QCtrl-Bench is a sixteen-task suite organized along three orthogonal axes of difficulty (system size, dynamical setting, and analytical tractability), whose per-axis task counts are summarized in Figure~\ref{fig:dataset}(e). Full per-task specifications are in Appendix~\ref{app:tasks}. Each task is defined by four components, together sufficient to drive the evaluation pipeline of Section~\ref{sec:pipeline} end-to-end:

(i) a \emph{natural-language specification}: the system Hamiltonian in standard operator notation with LaTeX markup for precise parsing, the initial and target states or unitaries, the control parameterization, and any physical constraints;

(ii) a \emph{control space}: the number of piecewise-constant time slices and total evolution time $T$, field bounds (e.g.\ phase angles $\alpha \in [-\pi,\pi]$, detuning $\nu \in [\nu_{\min}, \nu_{\max}]$, depending on the system), fixed physical parameters (Rabi frequency $\Omega$, coupling $J$, drift coefficients, energy gap $\Delta$), and the number and type of independent control channels;

(iii) a \emph{machine-verifiable fidelity objective} with target tolerance $\varepsilon$: either a state-overlap or a gate-overlap on a Hilbert space of dimension $d$,
\begin{equation}
\begin{aligned}
  F_{\mathrm{state}} &= |\langle \psi_{\mathrm{target}} | \psi(T) \rangle|^2, \\
  F_{\mathrm{gate}}(U_{\mathrm{target}}, U) &= \frac{1}{d^2}\,|\mathrm{Tr}(U_{\mathrm{target}}^\dagger U)|^2;
\end{aligned}
\label{eq:fidelity_conventions}
\end{equation}
and

(iv) a \emph{classical baseline fidelity}: a precomputed fidelity from GRAPE (numerical-only tasks), CRAB (frequency-tuned tasks), or the closed-form protocol (the seven analytically tractable tasks), used only as a comparison target in the Results. For the seven analytically tractable tasks, the closed-form protocol additionally serves as the interpretability anchor of Section~\ref{sec:interp} and is never exposed to the LLM. Scoring is defined entirely by (iii); the LLM sees only (i) and (ii), plus a zero-amplitude warm-start with its computed fidelity at iteration $k=1$ for VibeFn-QCtrl (Section~\ref{sec:fspsa}).

\emph{Calibration noise.} Each task is evaluated under two conditions: a noiseless baseline ($\sigma = 0$) and a miscalibrated drift Hamiltonian ($\sigma = 0.02$). In the miscalibrated condition, at the start of each outer-loop run every drift parameter $p$ receives a fixed multiplicative offset $p \to p(1+\delta_p)$ with $\delta_p \in \{-\sigma, +\sigma\}$ chosen once, independently per parameter, and held constant for the entire coefficient-optimization loop and for every subsequent evaluation within that run (static miscalibration, not shot noise). The LLM sees only the nominal parameters, so the $\sigma = 0.02$ condition tests whether a protocol remains high-fidelity when the true Hamiltonian differs from the one it was designed against---the regime in which real hardware always operates.

\subsection{Baselines}
\label{sec:baselines}

We compare VibeFn-QCtrl against two structurally different families of alternative. \emph{Classical numerical controllers} (GRAPE, CRAB, deep-learning optimizers, and gradient-free pulse-amplitude optimization) act as an \emph{external reference}: they fix the floor and ceiling of attainable fidelity on each task and are precomputed once with the benchmark. \emph{Prompt-only LLM baselines} (Simple LLM and Hint LLM) act as an \emph{ablation of the form-proposal step}: they share VibeFn-QCtrl's outer-loop scaffolding but the model must emit numerical amplitudes directly, bypassing form proposal. Both families are run under the same noise model and against the same fidelity objective as VibeFn-QCtrl, so any gap is attributable to the method rather than to the evaluation protocol.

For each task, the classical baseline is derived from the chosen optimizer used in the prior literature that originally introduced or studied that task. Under zero noise, all 16 tasks in \ref{app:tasks} uses one of the three choices: GRAPE~\cite{khaneja2005} (gradient-based, L-BFGS-B) for numerical-only tasks, CRAB~\cite{doria2011} (Fourier-basis, Nelder--Mead) for frequency-tuned tasks, and the closed-form analytical protocol for analytically tractable tasks. Under noise level, $\sigma = 0.02$, the classical baseine is a Chopped Random Basis (CRAB) pulse optimized via the SPSA algorithm (CRAB+SPSA), which is chosen as a standard robust gradient-free method for pulse optimization under calibration noise. Specifically, the control field for each channel $c$ is parameterized multiplicatively as $u_c(t) = u_{0,c}(t) (1 + f_c(t))$, where $u_{0,c}(t)$ is the initial control guess (either a random initialization or a GRAPE solution of the noise-free problem), and $f_c(t) = \sum_{k=1}^m [ A_{c,k} \sin(\omega_{c,k} t) + B_{c,k} \cos(\omega_{c,k} t) ]$ (where $m=20$) is a chopped random basis modulation signal with randomized frequencies $\omega_{c,k}$ and optimized coefficients $\{A_{c,k}, B_{c,k}\}$. SPSA (following the same update rule of Eq.~\eqref{eq:spsa}) is then applied directly to optimize the coefficients $\{A_{c,k}, B_{c,k}\}$ under calibration noise, exposed to the same noise as VibeFn-QCtrl.

Both pure-LLM baselines share VibeFn-QCtrl's outer-loop budget $K$ but remove the coefficient optimizer entirely: the model must emit numerical amplitudes directly. The two conditions differ in both the physics hint and whether structured feedback is returned between iterations:
\begin{itemize}
  \item \emph{Simple LLM}: a minimal hint consisting of the single injunction ``Think!!'', with no physics context beyond the task specification and no reflection-agent feedback between iterations.
  \item \emph{Hint LLM}: a physics-aware hint that names relevant control paradigms and lists typical noise sources. The full text of the hint is reproduced verbatim in Appendix~\ref{app:hint_prompt}. The reflection-agent feedback block carried from previous iterations.
\end{itemize}
In both baselines, the LLM writes the pulse out directly as a list of numerical amplitudes rather than as an analytic formula in VibeFn-QCtrl methods. The fidelity is calculated from the proposed optimized coefficients.

\subsection{Evaluation pipeline and metrics}
\label{sec:pipeline}

Each \emph{control task} runs through a four-phase pipeline for LLM approaches(Figure~\ref{fig:dataset}(d)):
\begin{itemize}
  \item \emph{Prompt assembly}: packages the task specification (Hamiltonian, initial and target states, control parameterization, constraints) into a single prompt. For VibeFn-QCtrl and Hint LLM, a rolling memory of previous rounds (prior proposals and reflection notes) is also appended.
  \item \emph{Pulse proposal}: sends the prompt to the LLM and parses its reply. VibeFn-QCtrl receives a short analytic expression per control channel (for example, $f(t)=\Omega\sin(\omega t)$ with a few free coefficients), which the coefficient optimizer then tunes against the simulator for up to $B_{\mathrm{opt}} = 10^3$ steps; Simple LLM and Hint LLM receive a list of pulse amplitudes, one per time slice, taken as-is.
  \item \emph{Fidelity evaluation}: computes the quantum dynamics under the (possibly noisy) controls and computes the task fidelity.
  \item \emph{Reflection}: closes the round. For VibeFn-QCtrl and Hint LLM, a separate reflection agent summarizes the round into bullets that are appended to the next round's prompt; Simple LLM has no reflection information and starts each round from the unchanged task prompt.
\end{itemize}
Rounds repeat until $1 - F^\star \le \varepsilon$ or the outer-loop budget $K = 50$ is reached. Classical baselines are precomputed once per task outside this pipeline (Section~\ref{sec:baselines}). The pipeline is built on the SDE-Harness~\cite{song2025sde} workflow engine with a LiteLLM~\cite{litellm2023} provider layer.

Four frontier LLMs are evaluated across all applicable methods: Claude Sonnet 4.5, Gemini 3.0 Flash, GPT-OSS-120B, and Qwen 3.5 Plus. We report three metrics: the \emph{task fidelity} $F$; the \emph{token efficiency} (output tokens per repetition); and the \emph{iteration efficiency} (outer-loop iterations to reach tolerance, capped at $K = 50$).

We use $n = 1$ repetition for VibeFn-QCtrl and $n = 5$ repetitions for Simple LLM and Hint LLM (reporting the best repetition over the five). The corresponding per-model fidelity distributions are reported in Figure~\ref{fig:prompt_barplot_noise002} ($\sigma = 0.02$, main text) and Figure~\ref{fig:prompt_fidelity_combined} ($\sigma = 0$, Appendix).

\subsection{Scaling-study protocols}
\label{sec:scaling-detail}

The two scaling control tasks in Figure~\ref{fig:scaling_law_combined} use sampling protocols that differ from the main-text benchmark. All scaling experiments use Claude Sonnet 4.5 as the underlying LLM. For the inference-time scaling panels (a,\,b), each task runs for $n = 80$ independent VibeFn-QCtrl repetitions with distinct random seeds, in contrast to the main-text per-task convention of $n = 1$ VibeFn-QCtrl repetition. For the pulse-resolution scaling panels (c,\,d), each task uses $n = 1$ VibeFn-QCtrl repetition and $n = 5$ repetitions per pure-LLM baseline, but at four pulse-resolution settings $N \in \{20, 80, 160, 220\}$.

\section*{Data availability}
The data that support the findings of this study are available from the corresponding author upon reasonable request.

\section*{Acknowledgments}

JH acknowledges support from the Natural Science Foundation of Jiangsu Province (No. BK20250404), the Youth Science and Technology Talent Support Project of Jiangsu Province (No. JSTJ-2025-600). DL acknowledges support from the Beijing Major Science and Technology Project under Contract No. Z251100008125036. 

\section*{Competing interests}
\begingroup
The authors declare no competing interests.
\endgroup

\clearpage
\appendix

\section{Benchmark Task Suite}
\label{app:tasks}

QCtrl-Bench comprises sixteen quantum control tasks, each probing a distinct aspect of control synthesis. The tasks are presented below in their canonical ordering, with the underlying physics and control objective specified for each.

\subsection{Task I: Qubit Ground-State Transfer}
\label{sec:counterdiabatic}

Counter-diabatic (CD) driving, also known as shortcuts to adiabaticity, enables high-fidelity quantum state transfer through an auxiliary control field that cancels diabatic transitions~\cite{demirplak2003, berry2009}. This task extends CD driving to time-dependent parameters using higher-order polynomial ramps to ensure smoothness.

\paragraph{Hamiltonian.} The system Hamiltonian is defined in terms of a time-dependent tunneling amplitude $\Delta(t)$ and a longitudinal field $\nu(t)$:
\begin{equation}
    H(t) = \Delta(t)\sigma^x + \nu(t)\sigma^z - g(t)\sigma^y,
\end{equation}
where $g(t)$ is the CD control field to be optimized. The system parameters evolve according to
\begin{equation}
\begin{aligned}
    \Delta(t) &= \Delta_0\left[1+\left(\frac{t}{T}\right)^3\right], \\
    \nu(t) &= h_0 + (h_f - h_0)\,S\!\left(\frac{t}{T}\right),
\end{aligned}
\end{equation}
where $S(\tau) = 10\tau^3 - 15\tau^4 + 6\tau^5$ (with $\tau = t/T$) is the quintic smoothstep function.

\paragraph{Exact solution.} The CD field that ensures perfect adiabatic following is given by
\begin{equation}
    g_{\mathrm{exact}}(t) = \frac{\Delta(t)\dot{\nu}(t) - \nu(t)\dot{\Delta}(t)}{2\left(\Delta(t)^2 + \nu(t)^2\right)},
\end{equation}
with $\dot{\Delta}(t) = 3\Delta_0 t^2/T^3$ and $\dot{\nu}(t) = (h_f - h_0)\,(30/T)\,\tau^2(1-\tau)^2$ (where $\tau = t/T$).

\paragraph{Calibration noise.} To model hardware miscalibration, multiplicative scaling is applied to all Hamiltonian parameters,
\begin{equation}
\begin{aligned}
    \Delta_{0,\mathrm{noisy}} &= \Delta_0(1+\delta), \\
    h_{0,\mathrm{noisy}} &= h_0(1+\delta), \\
    h_{f,\mathrm{noisy}} &= h_f(1+\delta),
\end{aligned}
\end{equation}
and the resulting physical fidelity is reported at the representative error level $\delta = \pm 0.02$. Because global parameter scaling alters the required cancellation geometry relative to the fixed control field $g(t)$, this perturbation provides a stringent robustness test of the protocol.

\subsection{Task II: Dicke State Preparation}
\label{sec:dicke}

Dicke states are symmetric multi-qubit entangled states with fixed excitation number~\cite{dicke1954}. The Dicke state $|D_{N_q}^{(k)}\rangle$ is the equal superposition of all $N_q$-qubit computational basis states of Hamming weight exactly $k$:
\begin{equation}
    |D_{N_q}^{(k)}\rangle = \binom{N_q}{k}^{-1/2} \sum_{|\mathbf{x}|=k} |\mathbf{x}\rangle,
\end{equation}
where the sum runs over all $N_q$-bit strings $\mathbf{x}$ of Hamming weight $k$.

To make the underlying symmetry manifest and to reduce the effective dimension, we replace the full $N_q$-qubit computational basis with the symmetric collective-spin basis. The collective spin operators are defined as
\begin{equation}
    J_{\mu} = \tfrac{1}{2}\sum_{i=1}^{N_q} \sigma_{\mu}^{(i)},\qquad \mu\in\{x,y,z\},
\end{equation}
where $\sigma_{\mu}^{(i)}$ is the Pauli-$\mu$ operator acting on qubit $i$, together with the total-spin operator $J^2 = J_x^2 + J_y^2 + J_z^2$. The fully symmetric subspace carries total spin $j=N_q/2$ and admits Dicke basis states $|j,m\rangle$ satisfying
\begin{equation}
    J^2|j,m\rangle = j(j+1)|j,m\rangle,\qquad J_z|j,m\rangle = m|j,m\rangle,
\end{equation}
with $m\in\{-j,\ldots,j\}$~\cite{yang2025qcircuitbench}. Here $J_z$ is the collective spin projection along the $z$ axis (equivalently, the sum of the individual $z$ components), and $m$ is related to the excitation number $k$ by $m=k-N_q/2$. The symmetric weight-$k$ superposition $|D_{N_q}^{(k)}\rangle$ thus coincides with $|j,m\rangle$, and the all-zero state satisfies $|0\cdots 0\rangle\equiv |j,-j\rangle$.

We consider a collective-spin Hamiltonian with two phase controls,
\begin{equation}
    H(x_1, x_2) = \omega\big[\cos(x_1) J_x + \cos(x_2) J_y\big] + \beta J_z^2,
\end{equation}
where $J_x,J_y,J_z$ act on the symmetric subspace. The controls are piecewise-constant phase sequences $x_1[t],x_2[t]\in[-\pi,\pi]$ discretized into $N$ time slices of duration $\Delta t = T/N_t$. The resulting evolution is
\begin{equation}
\begin{aligned}
    U(T) &= \prod_{i=1}^{N} \exp\bigl(-i H(x_1^i,x_2^i)\,\Delta t\bigr), \\
    |\psi(T)\rangle &= U(T)|\psi(0)\rangle,
\end{aligned}
\end{equation}
and solutions are evaluated by the state fidelity $F = |\langle D_{N_q}^{(k)}|\psi(T)\rangle|^2$.

\paragraph{GRAPE optimization.} As a classical baseline, the piecewise-constant phases are optimized via a GRAPE-style discretization: the loss $\mathcal{J}=1-F$ is minimized over the $2N_t$ decision variables $\{x_1^i,x_2^i\}_{i=1}^{N_t}$ using a quasi-Newton method such as L-BFGS-B~\cite{khaneja2005}. Robustness to calibration errors is additionally probed by evaluating the resulting pulse under multiplicative miscalibration of $(\omega,\beta)$.

\paragraph{Calibration noise.} Hardware miscalibration is modeled by multiplicative scaling of the Hamiltonian parameters,
\begin{equation}
    \omega_{\mathrm{noisy}} = \omega(1+\delta),\qquad \beta_{\mathrm{noisy}} = \beta(1+\delta),
\end{equation}
and the resulting physical fidelity is reported at the representative error level $\delta=\pm 0.02$. This stress-tests whether proposed pulses remain effective when both the transverse driving and the interaction strength drift together.

\subsection{Task III: Avoided-Crossing State Transfer}
\label{sec:landauzener}

The Landau--Zener model~\cite{landau1932, zener1932} considers a two-level system (qubit) governed by the time-dependent Hamiltonian
\begin{equation}
    H(t) = \Delta\,\sigma_x + \nu(t)\,\sigma_z,
\end{equation}
where $\sigma_x$ and $\sigma_z$ are Pauli matrices, $\Delta$ is a constant transverse tunnel coupling, and $\nu(t)$ is the time-dependent longitudinal detuning. The instantaneous eigenenergies are $E_\pm(\nu) = \pm\sqrt{\Delta^2 + \nu^2}$, with corresponding excited-state eigenvector
\begin{equation}
    |\phi(\nu)\rangle = \cos(\theta/2)|0\rangle + \sin(\theta/2)|1\rangle, \qquad \tan\theta = \Delta/\nu.
\end{equation}

In the present numerical studies we fix $\Delta = 1.2$ and a total sweep range $\nu(t) \in [-\nu_{\mathrm{limit}}, +\nu_{\mathrm{limit}}]$ with $\nu_{\mathrm{limit}} = 3$. The objective is to obtain from the LLM an optimal control protocol $\nu(t)$ that drives the system from the initial state $|\psi_0\rangle = |\phi(-\nu_0)\rangle$ to the target state $|\psi_*\rangle = |\phi(+\nu_0)\rangle$ within a fixed duration $T = \pi/2$, with $\nu_0 = 3$.

The control field $\nu(t)$ is discretized into $N_t=20$ piecewise-constant time slices of duration $\delta t = T/N_t$. The total unitary evolution operator is approximated by the product of discrete propagators,
\begin{equation}
    U(T) = \prod_{k=1}^{N} \exp\!\left[ -i\bigl(\Delta\,\sigma_x + \nu_k\,\sigma_z\bigr)\,\delta t \right],
\end{equation}
where $\nu_k$ is the constant amplitude on the $k$-th interval. Under VibeFn-QCtrl, the LLM emits a continuous function $f(t)$, which is sampled at the slice midpoints $t_k = (k+\tfrac{1}{2})\,\delta t$ to obtain $\nu_k = f(t_k)$. Performance of the resulting sequence $\{\nu_k\}$ is quantified by the state fidelity
\begin{equation}
    \mathcal{F} = |\langle\psi_*| U(T) |\psi_0\rangle|^2.
\end{equation}

If any $\nu_k$ supplied by the LLM falls outside $[-\nu_{\mathrm{limit}}, +\nu_{\mathrm{limit}}]$, the evaluator returns $\mathcal{F}=0$ as a penalty; this rule is disclosed to the LLM in the prompt.

The task evaluates the ability of the LLM to maximize the final-state fidelity $\mathcal{F}$. We benchmark a range of LLMs and prompt strategies against classical numerical methods including GRAPE (Gradient Ascent Pulse Engineering) and a gradient-free CRAB (Chopped Random Basis) optimizer~\cite{doria2011}. To assess robustness, evaluations are conducted across several noise levels: a noise level of $M\%$ is implemented by scaling the transverse coupling $\Delta$ by a factor of $(1 + M\%)$. The actual value of $\Delta$ is withheld from the LLM during optimization, so that the model must discover effective control sequences through a purely black-box iterative feedback loop.

\subsection{Task IV: Phase-Modulated Qubit Rotation}
\label{sec:squnitary}

Unitary synthesis generalizes state preparation: rather than mapping a single initial state to a single target, the goal is to implement an entire unitary transformation $U_{\mathrm{target}}\in \mathrm{SU}(2)$ over a fixed duration $T$ using phase-only control.

We consider a resonantly driven two-level system in the rotating frame (within the rotating-wave approximation) with a fixed drive amplitude and phase-only control~\cite{glaser2015}. The qubit evolves under the time-dependent Hamiltonian
\begin{equation}
    H(t) = \frac{\Omega}{2}\left[\cos\alpha(t)\,\sigma_x + \sin\alpha(t)\,\sigma_y\right],
\end{equation}
where $\sigma_x,\sigma_y$ are Pauli operators, $\Omega$ is the (constant) Rabi frequency---the angular rate of coherent rotations under resonant driving---and $\alpha(t)$ is the control phase, which selects the instantaneous rotation axis within the equatorial ($x$--$y$) plane of the Bloch sphere ($\alpha=0$ corresponds to an $x$-axis drive and $\alpha=\pi/2$ to a $y$-axis drive). The closed-system propagator satisfies the Schr\"odinger equation
\begin{equation}
    \dot U(t) = -i H(t) U(t),\qquad U(0)=\mathbb{I}.
\end{equation}

Following a GRAPE-style discretization, the control is parameterized as $N_t$ piecewise-constant segments over the fixed duration $T$: with $\Delta t = T/N_t$ and $\alpha(t)=\alpha_k$ for $t\in[k\Delta t,(k+1)\Delta t)$, the final unitary is
\begin{equation}
    U(T) = \prod_{k=0}^{N_t-1} \exp\!\big(-i\,H(\alpha_k)\,\Delta t\big),\qquad \alpha_k\in[-\pi,\pi].
\end{equation}

Solutions are evaluated using the trace-overlap gate fidelity $F_{\mathrm{gate}}(U_{\mathrm{target}}, U(T))$ of Eq.~\eqref{eq:fidelity_conventions} with $d=2$, which is invariant under an overall (global) phase of $U(T)$. Because $H(t)$ generates only equatorial rotations, the target gate is generically misaligned with the instantaneous control axis; universal single-qubit control therefore requires time-varying phase patterns that compose multiple non-commuting rotations. In particular, composite phase modulation can synthesize effective $z$-axis rotations such as $U_z(\pi)=\exp(-i\sigma_z\pi/2)$ despite the absence of a direct $\sigma_z$ control term. This task tests whether the LLM can reason about universal control from constrained Hamiltonians and produce phase sequences that realize a full target unitary rather than a single state transfer.

\paragraph{GRAPE optimization.} As a classical baseline, the phase sequence $\{\alpha_k\}_{k=0}^{N_t-1}$ is optimized under the same discretization by minimizing $\mathcal{J}=1-F_{\mathrm{gate}}$ with GRAPE-style gradients (e.g., using a quasi-Newton method such as L-BFGS-B).

\paragraph{Calibration noise.} Drive miscalibration is modeled by multiplicative scaling of the Rabi coefficient,
\begin{equation}
    \Omega_{\mathrm{noisy}} = \Omega(1+\delta),
\end{equation}
and the resulting physical performance is reported at the representative error level $\delta=\pm 0.02$.

\subsection{Task V: Two-Qubit State Transfer}
\label{sec:crab2qubit}

This task targets the transfer from an initial product state $|\psi_0\rangle = |00\rangle$ to the target $|\psi_{\mathrm{targ}}\rangle = |11\rangle$ through a controllable inter-qubit coupling. The system evolves under an Ising-type Hamiltonian with fixed random drift coefficients and a single interaction control:
\begin{equation}
    H(t) = \sum_{i=1}^2 \left(\alpha_i \sigma_x^{(i)} + \beta_i \sigma_z^{(i)}\right) + u(t)\, \sigma_z^{(1)}\sigma_z^{(2)},
\end{equation}
where $\alpha_i, \beta_i \in [0,1]$ are fixed random drift coefficients drawn from a predetermined seed (ensuring reproducibility across evaluations), and $u(t)$ is a piecewise-constant control field that modulates the Ising interaction strength~\cite{khaneja2005}. The state-transfer fidelity is $F = |\langle\psi_{\mathrm{targ}}|\psi(T)\rangle|^2$.

The control is discretized into $N_t=25$ time slices with total evolution time $T=18$. This task probes the ability of the LLM to reason about entanglement generation through controllable interactions and to navigate optimization landscapes induced by random drift Hamiltonians that break rotational symmetries.

\subsection{Task VI: Controlled-Phase Gate Synthesis}
\label{sec:cphase}

The controlled-phase (CPHASE) gate is a fundamental two-qubit entangling operation that applies a phase $\phi$ to the $|11\rangle$ state~\cite{khaneja2005}:
\begin{equation}
    U_{\mathrm{CPHASE}}(\phi) = \mathrm{diag}(1, 1, 1, e^{i\phi}),
\end{equation}
where $\phi=\pi$ yields the controlled-$Z$ gate. The system Hamiltonian includes a calibration-induced drift term and seven control channels:
\begin{equation}
\begin{aligned}
    H(t) &= (\kappa - 1)\,\sigma_x^{(1)}\sigma_x^{(2)} + \sum_{i=1}^2 \vec{u}_i(t)\cdot\vec{\sigma}^{(i)} \\
    &\quad + u_{\mathrm{db}}(t)\sum_{\mu\in\{x,y,z\}}\sigma_\mu^{(1)}\sigma_\mu^{(2)},
\end{aligned}
\end{equation}
where $\vec{u}_i(t)=(u_{ix},u_{iy},u_{iz})$ and $\vec{\sigma}=(\sigma_x,\sigma_y,\sigma_z)$. Here $\kappa$ is a calibration parameter (nominally $\kappa=1$, so that the drift term vanishes in the noiseless case), and $u_{\mathrm{db}}(t)$ is a single interaction channel. The gate fidelity is evaluated using the trace-overlap measure $F_{\mathrm{gate}}(U_{\mathrm{target}}, U(T))$ of Eq.~\eqref{eq:fidelity_conventions} with $d=4$.

The control is discretized into $N_t=25$ time slices with total evolution time $T=2\pi$. This task probes the ability of the LLM to synthesize high-fidelity two-qubit gates through coordinated multi-channel control.

\subsection{Task VII: Damped-Qubit Hadamard Gate}
\label{sec:lindbladian}

Open quantum systems require control design in the presence of dissipation and decoherence. This task targets implementation of the Hadamard gate $H = \tfrac{1}{\sqrt{2}}(\sigma_x + \sigma_z)$ on a single qubit subject to amplitude damping, described by the Lindblad master equation~\cite{lindblad1976, gorini1976}
\begin{equation}
    \frac{d\rho}{dt} = \mathcal{L}[\rho] = -i[H(t), \rho] + \gamma\left(\sigma_-\rho\sigma_+ - \tfrac{1}{2}\{\sigma_+\sigma_-, \rho\}\right),
\end{equation}
where the Hamiltonian is $H(t) = \tfrac{\omega_q}{2}\sigma_z + \tfrac{\Delta}{2}\sigma_x + u_z(t)\sigma_z + u_x(t)\sigma_x$ with $\omega_q=1.1$, $\Delta=0.15$, and damping rate $\gamma=0.15$. The controls $u_z(t)$ and $u_x(t)$ are piecewise-constant fields discretized into $N_t=10$ time slices with total evolution time $T=2$.

The evolution is computed in Liouville space as a superoperator: the initial map is $E_0 = \mathcal{I} \otimes \mathcal{I}$ (the identity superoperator) and the target is $E_{\mathrm{targ}} = H \otimes H^*$. The fidelity is measured by the trace-difference norm,
\begin{equation}
    F = 1 - \frac{1}{2d^2}\mathrm{Tr}\!\left[(E_{\mathrm{targ}} - E)^\dagger(E_{\mathrm{targ}} - E)\right],
\end{equation}
where $d=2$ is the Hilbert-space dimension. This task evaluates the ability of the LLM to reason about non-unitary dynamics, to design robust controls that compensate for dissipation, and to operate on superoperator representations of quantum channels.

\subsection{Task VIII: Two-Qubit Fourier Gate Synthesis}
\label{sec:qft_optimization}

We synthesize a two-qubit Quantum Fourier Transform (QFT) gate~\cite{coppersmith2002} using an optimized pulse parameterization. The physical system consists of two qubits coupled via an isotropic Heisenberg interaction with fixed coupling strength $J = 0.618$, described by the drift Hamiltonian
\begin{equation}
    H_{\mathrm{drift}} = J \left( \sigma_x^{(1)}\sigma_x^{(2)} + \sigma_y^{(1)}\sigma_y^{(2)} + \sigma_z^{(1)}\sigma_z^{(2)} \right).
\end{equation}
The control Hamiltonian provides independent driving of each qubit along the $x$ and $y$ axes, so that the total time-dependent Hamiltonian is
\begin{equation}
    H(t) = H_{\mathrm{drift}} + \sum_{q=1}^2 \left[ u_{x,q}(t)\sigma_x^{(q)} + u_{y,q}(t)\sigma_y^{(q)} \right],
\end{equation}
where the controls $u_{k,q}(t)$ are continuous functions parameterized by a truncated Fourier series. The objective is to maximize the gate fidelity $F_{\mathrm{gate}}(U_{\mathrm{target}}, U(T))$ of Eq.~\eqref{eq:fidelity_conventions} with $d=4$, with respect to the canonical two-qubit QFT unitary.

\subsection{Task IX: Dissipative Lambda Transfer}
\label{sec:lambda}

Population transfer in a three-level $\Lambda$-system is modeled by a non-Hermitian effective Hamiltonian that accounts for the lossy intermediate state $|2\rangle$ with decay rate $\gamma = 0.4$. In the rotating frame, with complex pump ($\Omega_P$) and Stokes ($\Omega_S$) fields, the Hamiltonian reads
\begin{equation}
    H_{\mathrm{eff}}(t) = \begin{pmatrix}
    \Delta_P & -\tfrac{1}{2}\Omega_P(t) & 0 \\
    -\tfrac{1}{2}\Omega_P^*(t) & -i\gamma & -\tfrac{1}{2}\Omega_S(t) \\
    0 & -\tfrac{1}{2}\Omega_S^*(t) & \Delta_S
    \end{pmatrix},
\end{equation}
where $\Delta_P$ and $\Delta_S$ denote the detunings of the pump and Stokes lasers, respectively. The objective is to transfer population efficiently from the initial state $|1\rangle$ to the target state $|3\rangle$, typically by invoking the Stimulated Raman Adiabatic Passage (STIRAP) mechanism~\cite{vitanov2017} so as to avoid populating the lossy intermediate state $|2\rangle$.

\subsection{Task X: Driftless Single-Qubit Gate}
\label{sec:single_qubit}

This task evaluates full controllability on the single-qubit Bloch sphere~\cite{dalessandro2007}. In contrast to settings with restricted interactions, the agent here has control over all three Pauli generators, with Hamiltonian
\begin{equation}
    H(t) = u_x(t)\sigma_x + u_y(t)\sigma_y + u_z(t)\sigma_z.
\end{equation}
The control inputs $u_x, u_y, u_z$ are piecewise-constant fields. For each problem instance, a random target unitary $U_{\mathrm{target}} \in \mathrm{SU}(2)$ is generated via Euler angles,
\begin{equation}
    U_{\mathrm{target}} = R_z(\phi)\, R_x(\theta) = e^{-i \phi \sigma_z/2}\, e^{-i \theta \sigma_x/2}.
\end{equation}
The objective is to synthesize $U_{\mathrm{target}}$ with high fidelity, thereby testing the ability to coordinate simultaneous rotations about non-commuting axes.

\subsection{Task XI: Coupled-Oscillator Symplectic Transform}
\label{sec:symplectic}

This task involves the control of two coupled harmonic oscillators in the continuous-variable regime~\cite{weedbrook2012}, described by the evolution of an element $S(t)$ in the symplectic group $\mathrm{Sp}(4, \mathbb{R})$. The dynamics are governed by
\begin{equation}
    \dot{S}(t) = \Omega\, H_{\mathrm{total}}(t)\, S(t),
\end{equation}
where $\Omega$ is the symplectic form and the generator is $G(t) = \Omega\,[A_0 + u(t)\,A_c]$. With phase-space vector $x=(q_1, p_1, q_2, p_2)^T$, the quadratic Hamiltonian is $H = \tfrac{1}{2}x^T A x$, and the drift and control matrices are
\begin{equation}
A_0=
\begin{pmatrix}
\omega_1 & 0 & g_1 & 0 \\
0 & \omega_1 & 0 & g_1 \\
g_1 & 0 & \omega_2 & 0 \\
0 & g_1 & 0 & \omega_2
\end{pmatrix},
\qquad
A_c=
\begin{pmatrix}
1 & 0 & 0 & 0 \\
0 & 1 & 0 & 0 \\
0 & 0 & 0 & 0 \\
0 & 0 & 0 & 0
\end{pmatrix},
\end{equation}
with $\omega_1 = \omega_2 = 1.414$ and $g_1 = 0.5$. The symplectic form is
\begin{equation}
    \Omega = \begin{pmatrix} 0 & I_2 \\ -I_2 & 0 \end{pmatrix}.
\end{equation}
The goal is to steer the system from the identity $I_4$ to a target symplectic rotation $S_{\mathrm{target}} = \exp(\Omega)$.

\subsection{Task XII: Toffoli Gate Synthesis}
\label{sec:toffoli}

The Toffoli (CCNOT) gate~\cite{toffoli1980} is optimized on a linear chain of three qubits. The system comprises a constant drift term describing the qubit frequencies and a set of controllable interaction terms,
\begin{equation}
    H(t) = H_0 + \sum_{k} c_k(t)\, H_k^{\mathrm{ctrl}},
\end{equation}
with drift Hamiltonian $H_0 = 2\pi \sum_{j=1}^3 \sigma_z^{(j)}$. The control set $\{H_k^{\mathrm{ctrl}}\}$ comprises $18$ operators covering both single-qubit rotations and pairwise interactions,
\begin{equation}
    \{H_k^{\mathrm{ctrl}}\} = \{ \sigma_\alpha^{(j)} \} \cup \{ \sigma_\alpha^{(i)}\sigma_\alpha^{(j)} \}, \qquad \alpha \in \{x,y,z\},
\end{equation}
where the pairwise terms act on the connected neighbors $(1,2)$ and $(2,3)$, with additional non-adjacent pairs included subject to connectivity. The objective is to realize the three-qubit Toffoli gate $U_{\mathrm{Toffoli}}$ at maximal process fidelity in the eight-dimensional Hilbert space.

\subsection{Task XIII: Transmon Logical $X$ Gate}
\label{sec:transmon_xgate}

This task optimizes a logical Pauli-$X$ gate on a multilevel transmon qubit~\cite{koch2007transmon, krantz2019} in the charge basis. With basis states $\{|j\rangle\}_{j=-N_c}^{N_c}$, the drift Hamiltonian is
\begin{equation}
\begin{aligned}
    H_0 &= 4E_C \sum_{j=-N_c}^{N_c} (j-n_g)^2 |j\rangle\langle j| \\
    &\quad - \frac{E_J}{2}\sum_{j=-N_c}^{N_c-1} \left( |j+1\rangle\langle j| + |j\rangle\langle j+1| \right).
\end{aligned}
\end{equation}
Control enters through the charge operator
\begin{equation}
    q = \sum_{j=-N_c}^{N_c} (-2j)\, |j\rangle\langle j|,
\end{equation}
so that the full Hamiltonian reads
\begin{equation}
    H(t) = H_0 + V(t)\, q,
\end{equation}
with a single real control channel $V(t)$. The benchmark physical parameters are
\begin{equation}
\begin{aligned}
    &N_c = 8, \qquad E_C = 0.386~\mathrm{GHz}, \\
    &E_J = 15.44~\mathrm{GHz}, \qquad n_g = 0,
\end{aligned}
\end{equation}
yielding a Hilbert-space dimension $2N_c+1 = 17$. The total gate time is $T = 10.0~\mathrm{ns}$, and the control is discretized into $N_t = 50$ piecewise-constant time slices.

The logical qubit is defined by the two lowest-energy eigenstates of $H_0$, denoted $|0_L\rangle$ and $|1_L\rangle$, and the target operation is a logical Pauli-$X$ on this subspace. Given the simulated unitary $U(T)$, the induced logical $2\times 2$ block is
\begin{equation}
    U_{\log} = \begin{pmatrix}
        \langle 0_L|U(T)|0_L\rangle & \langle 0_L|U(T)|1_L\rangle \\
        \langle 1_L|U(T)|0_L\rangle & \langle 1_L|U(T)|1_L\rangle
    \end{pmatrix},
\end{equation}
and the solution is evaluated by the logical gate fidelity $F_{\mathrm{gate}}(X, U_{\log})$ of Eq.~\eqref{eq:fidelity_conventions} with $d=2$, where $X = \bigl(\begin{smallmatrix} 0 & 1 \\ 1 & 0 \end{smallmatrix}\bigr)$. The optimization objective is therefore the minimization of $\mathcal{J} = 1 - F_{\mathrm{gate}}$.

\subsection{Task XIV: Leakage-Aware Transmon Excitation}
\label{sec:drag_pulse}

This task optimizes a leakage-suppressed $\pi$ pulse~\cite{motzoi2009} for a three-level transmon model with computational states $|0\rangle, |1\rangle$ and leakage state $|2\rangle$. The Hamiltonian is
\begin{equation}
\begin{aligned}
    H(t) &= \delta(t)\,|1\rangle\langle 1| + \bigl(2\delta(t)-\alpha\bigr)\,|2\rangle\langle 2| \\
    &\quad + \frac{\Omega_x(t)}{2}\, M_x + \frac{\Omega_y(t)}{2}\, M_y,
\end{aligned}
\end{equation}
with
\begin{equation}
\begin{aligned}
    M_x &= |0\rangle\langle 1| + \lambda\, |1\rangle\langle 2| + \mathrm{h.c.}, \\
    M_y &= -i\,|0\rangle\langle 1| - i\lambda\, |1\rangle\langle 2| + \mathrm{h.c.}
\end{aligned}
\end{equation}
Here $\Omega_x(t)$ is the in-phase drive, $\Omega_y(t)$ is the quadrature DRAG correction, and $\delta(t)$ is a detuning-compensation channel. The benchmark parameters are
\begin{equation}
\begin{aligned}
    &\alpha = 0.3~\mathrm{GHz}, \qquad \lambda = \sqrt{2} \approx 1.414214, \\
    &T = 60.0~\mathrm{ns}, \qquad N_t = 50,
\end{aligned}
\end{equation}
with piecewise-constant controls over $N_t$ equal time slices.

The system starts in the ground state $|\psi(0)\rangle = |0\rangle$, and the target is a $\pi$-pulse into $|\psi_{\mathrm{target}}\rangle = |1\rangle$. For a piecewise-constant control sequence, the final state is
\begin{equation}
    |\psi(T)\rangle = \prod_{k=1}^{N} \exp\!\bigl(-i H_k\, \Delta t\bigr)\,|0\rangle.
\end{equation}
The pulse is evaluated by the target-state fidelity $F = |\langle 1|\psi(T)\rangle|^2$, and the controls are optimized by minimizing the infidelity $\mathcal{J} = 1 - F$.

\paragraph{Calibration noise.} Hardware miscalibration is modeled by multiplicative scaling of the anharmonicity,
\begin{equation}
    \alpha_{\mathrm{noisy}} = \alpha(1+\delta_\kappa),
\end{equation}
and the resulting state fidelity is reported at the representative error level $\delta_\kappa = \pm 0.02$. This stress-tests whether the proposed pulses remain effective when the transmon's frequency structure drifts.

\subsection{Task XV: NMR Coherence Transfer}
\label{sec:two_spin_transfer}

This task considers coherence transfer in a weakly coupled heteronuclear two-spin NMR system. Let
\begin{equation}
    I_\mu = \frac{\sigma_\mu}{2} \otimes \mathbb{I}_2, \qquad S_\mu = \mathbb{I}_2 \otimes \frac{\sigma_\mu}{2}, \qquad \mu \in \{x,y,z\},
\end{equation}
where the first tensor factor acts on spin $I$ and the second on spin $S$. The time-dependent Hamiltonian is
\begin{equation}
\begin{aligned}
    H(t) &= 2\pi J\, I_z S_z + u_{Ix}(t)\, I_x + u_{Iy}(t)\, I_y \\
    &\quad + u_{Sx}(t)\, S_x + u_{Sy}(t)\, S_y,
\end{aligned}
\end{equation}
where $J$ is the scalar coupling constant and $u_{Ix},u_{Iy},u_{Sx},u_{Sy}$ are the four RF control channels. The benchmark parameters are
\begin{equation}
    J = 100~\mathrm{Hz}, \qquad T = 15.0~\mathrm{ms}, \qquad N_t = 60,
\end{equation}
so that the controls are discretized into $N$ piecewise-constant time slices of duration $\Delta t = T/N = 0.25~\mathrm{ms}$.

The initial and target operators are $\rho(0) = S_x$ and $\rho_{\mathrm{target}} = I_x$, respectively. For a piecewise-constant control sequence, the evolution is
\begin{equation}
    \rho(T) = U(T)\,\rho(0)\,U^\dagger(T), \qquad U(T) = \prod_{k=1}^{N}\exp\!\bigl(-i H_k\, \Delta t\bigr),
\end{equation}
where $H_k$ denotes the Hamiltonian on the $k$-th time slice. The optimization objective is to maximize the polarization-transfer efficiency
\begin{equation}
    F_{\mathrm{transfer}} = \mathrm{Tr}\!\bigl(I_x\, \rho(T)\bigr),
\end{equation}
equivalently to minimize the infidelity $\mathcal{J} = 1 - F_{\mathrm{transfer}}$. With $I_x = \sigma_x/2 \otimes \mathbb{I}_2$ normalized so that $\mathrm{Tr}(I_x^2) = 1$, $F_{\mathrm{transfer}}$ coincides with the state-overlap fidelity of Eq.~\eqref{eq:fidelity_conventions} whenever $\rho(T)$ is a pure state aligned with the target operator; it takes values in $[-1, +1]$ under unitary evolution and attains $F_{\mathrm{transfer}} = 1$ for perfect transfer.

\paragraph{Analytical solution.} The time-optimal transfer is achieved by the INEPT (Insensitive Nuclei Enhanced by Polarization Transfer) sequence~\cite{morris1979}, which consists of carefully timed $\pi/2$ pulses interleaved with free evolution under the $J$-coupling. In the product-operator formalism the theoretical pathway is
\begin{equation}
    S_x \xrightarrow{\tau = 1/(4J)} 2\, I_z S_y \xrightarrow{(\pi/2)_{Ix}} 2\, I_y S_y \xrightarrow{\tau = 1/(4J)} I_x,
\end{equation}
where $\tau = 1/(4J) = 2.5~\mathrm{ms}$ for $J = 100~\mathrm{Hz}$. Under ideal conditions this analytical solution attains $F_{\mathrm{transfer}} = 1$ (perfect transfer).

\paragraph{Calibration noise.} Hardware miscalibration is modeled by multiplicative scaling of the coupling constant,
\begin{equation}
    J_{\mathrm{noisy}} = J(1+\delta),
\end{equation}
and the resulting transfer efficiency is reported at the representative error level $\delta = \pm 0.02$. This stress-tests whether the proposed pulse sequences remain effective when the $J$-coupling strength deviates from its nominal value.

\subsection{Task XVI: Polynomial-Noise Qubit Refocusing}
\label{sec:decoherence_suppression}

This task inverts the usual gate-synthesis objective: the control is required to act trivially---the qubit should return to its initial state---despite a known, time-dependent dephasing term that would otherwise rotate it. The LLM must therefore design a control pulse whose effect exactly cancels the noise-induced dephasing over the evolution, a dynamical-decoupling problem.

\paragraph{Hamiltonian.} A single qubit evolves under
\begin{equation}
    H(t) = \frac{\beta(t)}{2}\,\sigma_z + \frac{\Omega(t)}{2}\,\sigma_x,
\end{equation}
where $\beta(t)$ is a deterministic $\sigma_z$ disturbance (a time-varying detuning) and $\Omega(t)$ is the $\sigma_x$ control to be optimized. The disturbance has polynomial structure,
\begin{equation}
    \beta(t) = \beta_0 + \beta_1\,t + \beta_2\,t^2,
\end{equation}
with default parameters $(\beta_0,\beta_1,\beta_2) = (0.5,\,2.0,\,20.0)$. The constant, linear, and quadratic components jointly probe whether the LLM can refocus non-stationary noise rather than only DC dephasing.

\paragraph{Control and evolution.} The control $\Omega(t)$ is discretized into $N_t = 100$ piecewise-constant segments over total evolution time $T = 1$. Slice $k$ samples $\beta$ at its midpoint $t_k = (k-\tfrac{1}{2})\,\Delta t$, with $\Delta t = T/N_t$, giving
\begin{equation}
    U(T) = \prod_{k=1}^{N} \exp\!\Big[-i\,H\!\big(\Omega_k,\,\beta(t_k)\big)\,\Delta t\Big].
\end{equation}

\paragraph{Objective.} The target operation is the identity, $U_{\mathrm{target}} = \mathbb{I}$. Performance is measured by the gate-overlap fidelity $F_{\mathrm{gate}}(\mathbb{I}, U(T))$ of Eq.~\eqref{eq:fidelity_conventions} with $d = 2$, which rewards any $\Omega(t)$ whose evolution, combined with the unavoidable $\sigma_z$ drive, returns the qubit to the identity up to a global phase. Standard analytical baselines in this regime are the Carr--Purcell and XY pulse trains~\cite{carr1954, viola1999}, which apply $\sigma_x$ $\pi$-pulses at times chosen so as to refocus the accumulated dephasing into a closed loop on the Bloch sphere.

\paragraph{Calibration noise.} Hardware miscalibration is modeled by a multiplicative scaling applied jointly to all noise parameters,
\begin{equation}
    \beta_{k,\mathrm{noisy}} = \beta_k(1+\delta), \qquad k \in \{0,1,2\},
\end{equation}
with $\delta = \pm 0.02$. The LLM is provided with the nominal coefficients, whereas the evaluator uses the perturbed values; this stress-tests whether the proposed pulses remain effective when the dephasing profile drifts simultaneously in amplitude and slope.

\section{Efficiency and per-model breakdowns}
\label{sec:efficiency}

\begin{figure}[!t]
  \centering
  \includegraphics[width=\columnwidth]{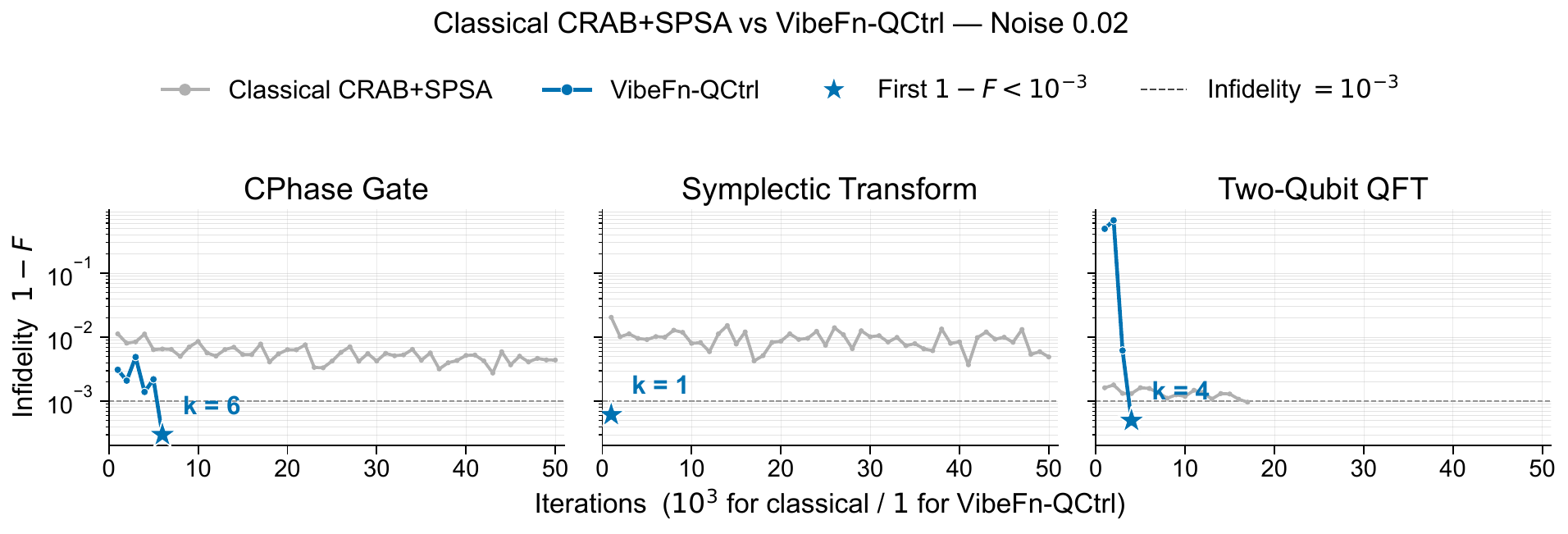}
  \caption{\textbf{Convergence trajectories at $\sigma = 0.02$ for three representative tasks.} Infidelity $1 - F$ (log scale) versus a shared iteration index for the classical baseline (grey) and VibeFn-QCtrl (blue) on (a) Controlled-Phase Gate Synthesis, (b) Coupled-Oscillator Symplectic Transform, and (c) Two-Qubit Fourier Gate Synthesis. One unit on the horizontal axis corresponds to $1000$ coefficient updates (CRAB+ SPSA for classical baseline; inner coefficient-optimizer updates for VibeFn-QCtrl), placing the two methods on matched budgets. The dashed horizontal line marks the threshold $1 - F = 10^{-3}$ ($F = 0.999$); blue stars mark the first VibeFn-QCtrl outer-loop iterations at which the threshold is crossed. The classical baseline does not cross the threshold within $50000$ iterations on (a) or (b), and first crosses it at iteration $17{,}000$ on (c).}
  \label{fig:efficiency}
\end{figure}

\subsection{Query efficiency vs.\ the classical baseline}

Beyond final fidelity, the cost to reach a target fidelity matters. We measure cost as the number of queries needed to first cross $F = 0.999$. For the classical baseline this is the SPSA iteration index, capped at $5 \times 10^4$. For VibeFn-QCtrl it is outer-loop iterations number $10^3$ (the per-round inner coefficient-optimizer budget).

Figure~\ref{fig:efficiency} shows convergence trajectories at $\sigma = 0.02$ for three representative tasks: Controlled-Phase Gate Synthesis, Coupled-Oscillator Symplectic Transform, and Two-Qubit Fourier Gate Synthesis. VibeFn-QCtrl crosses $F = 0.999$ within at most $6$ outer rounds ($6 \times 10^3$ simulator calls) on all three. The classical baseline does not cross the threshold within $5 \times 10^4$ iterations on Controlled-Phase Gate Synthesis or Coupled-Oscillator Symplectic Transform, and first crosses it only at iteration $17{,}000$ on Two-Qubit Fourier Gate Synthesis, between $4\times$ (Two-Qubit Fourier Gate Synthesis) and $\geq 50\times$ (Coupled-Oscillator Symplectic Transform) later in iteration number than VibeFn-QCtrl. The contrast reflects the dimensional-reduction mechanism of form proposal: once the LLM has committed to a compact analytic ansatz, the inner coefficient optimizer refines a handful of coefficients rather than tens of pulse amplitudes per channel.

\subsection{Token efficiency vs.\ pure-LLM baselines}

VibeFn-QCtrl token cost is comparable to or less than the pure-LLM methods across all evaluated LLMs and both noise levels. Per LLM, VibeFn-QCtrl ranges over roughly $0.4$--$1.05$ million tokens; Hint LLM and Simple LLM range over $0.4$--$1.45$ million. The asymmetry is largest for Qwen 3.5 Plus, where VibeFn-QCtrl uses $\sim\!0.5$--$0.85$M tokens versus $\sim\!1.4$M for the two pure-LLM methods.
\begin{figure}[H]
    \centering
    \includegraphics[width=\columnwidth]{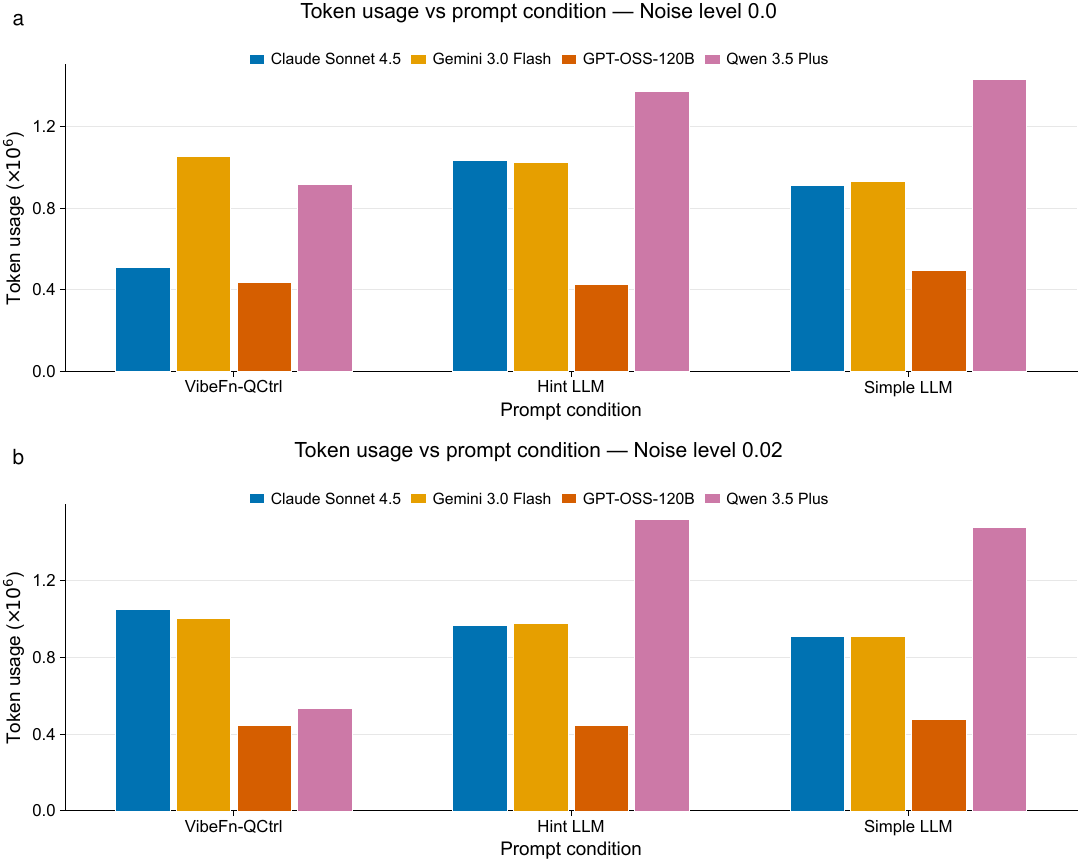}
    \caption{\textbf{Average total token consumption by method, model, and noise level.} Total tokens for the best run, averaged over the sixteen tasks. Methods on the x-axis are VibeFn-QCtrl, Hint LLM, and Simple LLM, coloured by LLM (Claude Sonnet 4.5, Gemini 3.0 Flash, GPT-OSS-120B, Qwen 3.5 Plus). \textbf{(a)}~$\sigma = 0$. \textbf{(b)}~$\sigma = 0.02$. For every LLM, VibeFn-QCtrl consumes comparable or fewer tokens than the two pure-LLM methods.}
    \label{fig:prompt_tokens_combined}
  \end{figure}

\subsection{Per-model fidelity at $\sigma = 0$}

Under VibeFn-QCtrl the four-model mean-fidelity spread is small compared to the pure-LLM methods. The pattern matches the $\sigma = 0.02$ result in main-text Figure~\ref{fig:prompt_barplot_noise002}: once the LLM has proposed the right functional form, the remaining coefficient optimization largely absorbs the model-capability gap.
\begin{figure}[H]
    \centering
    \includegraphics[width=\columnwidth]{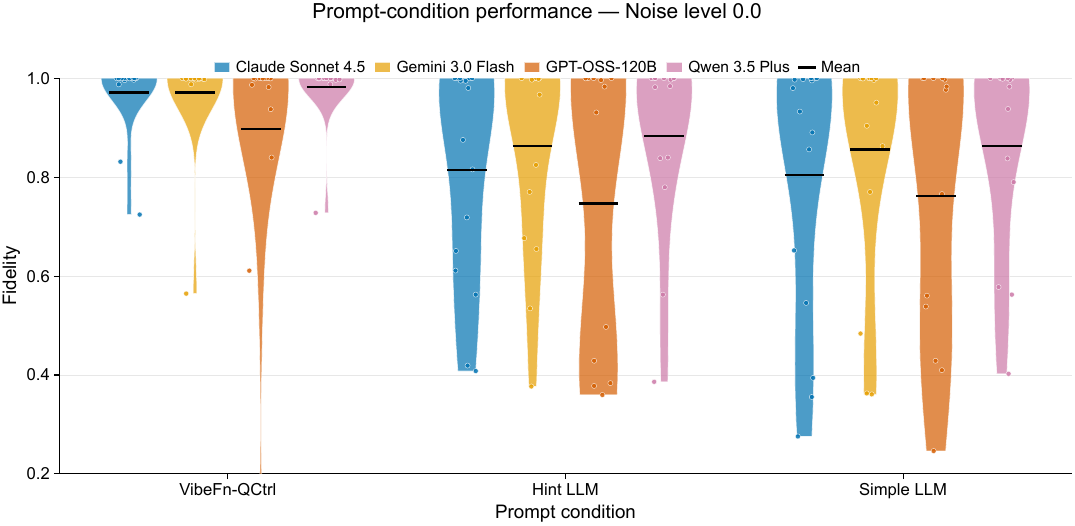}
    \caption{\textbf{Per-task best-fidelity distribution at $\sigma = 0$.} Violin plots of the per-task best fidelity over the sixteen QCtrl-Bench tasks, grouped by method (VibeFn-QCtrl, Hint LLM, Simple LLM) and coloured by LLM (Claude Sonnet 4.5, Gemini 3.0 Flash, GPT-OSS-120B, Qwen 3.5 Plus); the horizontal black tick on each violin marks the per-cell mean. The $\sigma = 0.02$ counterpart is shown in main-text Figure~\ref{fig:prompt_barplot_noise002}.}
    \label{fig:prompt_fidelity_combined}
  \end{figure}

\section{Iteration-prompt templates: rendered examples}
\label{app:prompt_templates}

This appendix shows the actual prompts the proposal agent receives at iteration $k\ge 2$, with every template variable substituted by its rendered value for the Dicke-state task (Appendix~\ref{app:tasks}, Task~II). The intent is to make concrete the three-block structure described in Methods, Section~\ref{sec:fspsa}: a task specification (Hamiltonian, controls, target, constraints, output schema), a physics-guidance block (the \texttt{hint\_prompt}), and a rolling feedback memory (the \texttt{previous\_repetitions}, the best-so-far metric, the reflection-agent \texttt{feedback} bullets, and a trailing instruction). 
\begin{figure*}[t!]
  \begin{center}
  \includegraphics[page=1, width=\textwidth, keepaspectratio]{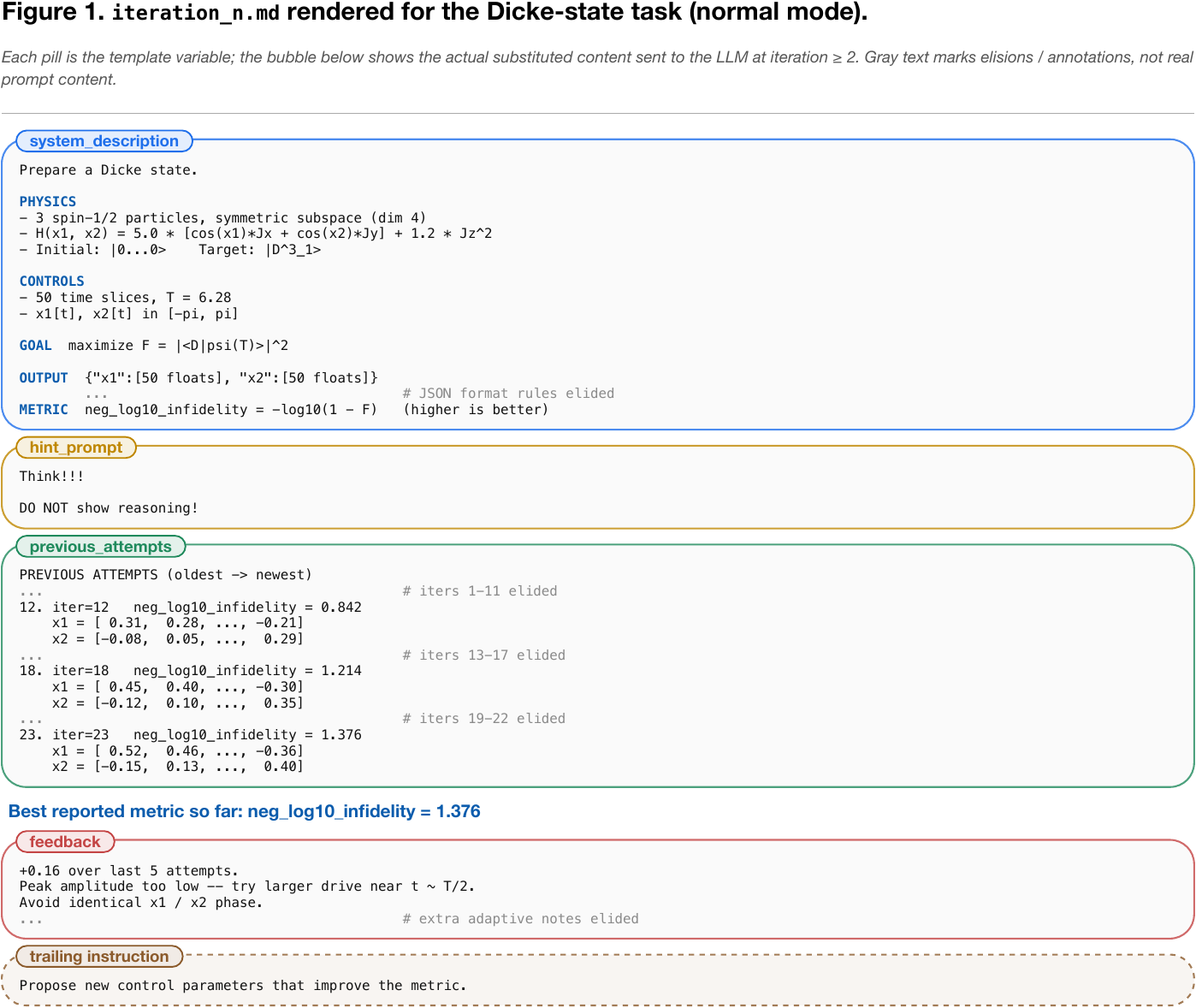}
  \end{center}
  \caption{\textbf{Rendered iteration prompt for the pure-LLM baselines (\texttt{iteration\_n.md}, numeric-controls mode), Dicke-state task.} Each colored pill is a template variable (\texttt{system\_description}, \texttt{hint\_prompt}, \texttt{previous\_repetitions}, \texttt{feedback}, trailing instruction); the bubble below shows the substituted content delivered to the LLM at iteration $k\ge 2$. In this mode the model must emit \texttt{x1[t]} and \texttt{x2[t]} as length-$50$ numeric arrays of pulse amplitudes; the rolling feedback memory stores those arrays verbatim. Gray text (``\dots'' lines, trailing ``\texttt{\#~\dots}'' annotations) marks editorial elisions only.}
  \label{fig:prompt_template_numeric}
\end{figure*}
The two figures correspond to the two methods that consume the template:

\begin{itemize}
  \item Figure~\ref{fig:prompt_template_numeric} renders \texttt{iteration\_n.md}, the \emph{numeric-controls} template used by the pure-LLM baselines (Simple LLM, Hint LLM). The model is asked to emit per-time-slice amplitude arrays \texttt{x1[t]} and \texttt{x2[t]} of length $50$, together with a free-form reflection-agent note carried forward as feedback.
  \item Figure~\ref{fig:prompt_template_function} renders \texttt{iteration\_n\_function\_class.md}, the \emph{function-class} template used by VibeFn-QCtrl. The system block now declares a parameterized symbolic grammar, the previous-repetitions memory stores analytic expressions instead of numeric arrays, and the trailing instruction steers the proposal agent to make the smallest structural change that improves the metric.
\end{itemize}

In both figures the colored pill above each bubble names the template variable; the bubble below shows its substituted content. Section headers inside the bubbles (\textsc{physics}, \textsc{controls}, \textsc{goal}, \dots) are highlighted in blue. Gray lines marked with ``\dots'' or trailing ``\texttt{\#~\dots}'' annotations are editorial elisions added for the figure only; they are \emph{not} part of the prompt text seen by the model. The full unelided templates ship with the open-source repository.

One directive in the \texttt{hint\_prompt} block of both figures warrants a brief note. The instruction ``\texttt{DO NOT show reasoning!}'' asks the model not to inline its chain-of-thought into the response body. Without it, some models inline long reasoning traces alongside the pulse output, which grows the context across iterations and crowds out the rolling feedback memory once the number of returned parameters is large; this is a practical context-window safeguard, not a constraint on the underlying reasoning step. On reasoning-capable models the chain-of-thought is still emitted through the provider's dedicated reasoning channel where one exists, and is not parsed into the simulator pipeline, so the directive controls placement only and does not affect the reported fidelity.

\begin{figure*}[t!]
  \begin{center}
  \includegraphics[page=2, width=\textwidth, keepaspectratio]{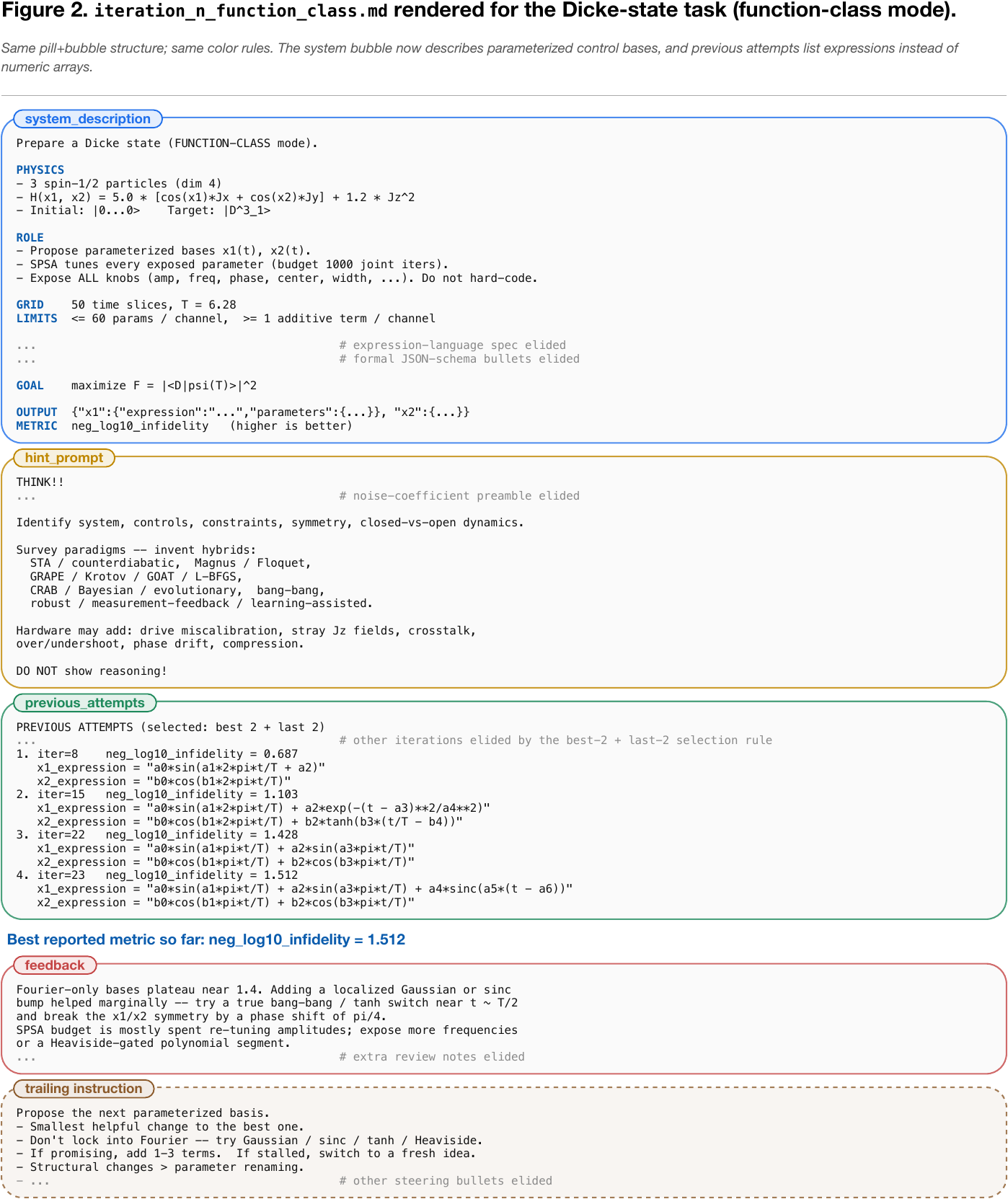}
  \end{center}
  \caption{\textbf{Rendered iteration prompt for VibeFn-QCtrl (\texttt{iteration\_n\_function\_class.md}, function-class mode), Dicke-state task.} Same pill-and-bubble structure as Figure~\ref{fig:prompt_template_numeric}, with three differences. (i) The \texttt{system\_description} now declares the parameterized symbolic grammar of Section~\ref{sec:fspsa} (at most $60$ free coefficients per channel, at least one additive term, allowed operations and functions); the model must return analytic expressions \texttt{x1.expression} and \texttt{x2.expression} with named parameters, which the inner SPSA optimizer then refines under a fixed budget of $10^3$ joint iterations. (ii) The \texttt{previous\_repetitions} memory stores the best two and the most recent two analytic ans\"atze together with their optimized infidelities, not numeric pulse arrays. (iii) The trailing instruction steers the proposal agent toward the smallest structural change to the best ansatz so far, favoring new basis families (Gaussian, sinc, tanh, Heaviside) over parameter renaming. Color and elision conventions match Figure~\ref{fig:prompt_template_numeric}.}
  \label{fig:prompt_template_function}
\end{figure*}

\subsection{Full text of the Hint LLM physics-guidance prompt}
\label{app:hint_prompt}

The block below reproduces verbatim the content of the \texttt{hint\_prompt} template used by the Hint LLM baseline (Methods, Section~\ref{sec:baselines}). It is appended to every iteration prompt the Hint LLM receives, between the task specification and the rolling feedback memory. The Simple LLM baseline replaces this entire block with the single line ``\texttt{Think!!}'' and is therefore not reproduced separately.

\begin{quote}\small
\noindent\texttt{THINK!!}

\noindent Realizing that the coefficient in this simulation might be affected by experimental condition.

\noindent Identify the system, available controls, constraints, and objective, and assess symmetry, controllability, and whether the dynamics are closed or open.

\noindent Actively survey both analytic and numerical paradigms: adiabatic and nonadiabatic control (STA, counterdiabatic driving, holonomic/geometric control), average-Hamiltonian/Magnus/Floquet engineering, invariant-based and time-optimal/bang-bang control, gradient based control (GRAPE, Krotov, GOAT, L-BFGS), gradient free optimization (CRAB, annealing, evolutionary algorithm, bayesian optimization), as well as robust control, measurement feedback, and learning-assisted methods.

\noindent Do not treat this list as exhaustive: prioritize achieving the best results by inventing hybrid or novel strategies that exploit physical structure, hidden noise effects and any feedback data, then select and execute the best-performing approach consistent with the objective and constraints.

\noindent Note: Due to systematic hardware imperfections and environmental noise, the effective physical realization may deviate from the intended model. Potential sources of deviation include: Drive Strength Miscalibration, Stray Longitudinal Fields, Control Crosstalk, Pulse Overshoot/Undershoot, Phase Reference Drift, and Nonlinear Compression.

\noindent\texttt{DO NOT show reasoning!}
\end{quote}

\begin{widetext}
\section{LLM Reasoning Traces for Zero-Step Coefficient-Optimizer Tasks}
\label{app:reasoning_traces}

This appendix provides the detailed reasoning traces generated by the Qwen 3.5 Plus model under the zero-coefficient-optimizer setting (SPSA iteration = 0) for the three analytically structured tasks discussed in Section~\ref{sec:interp}. In this zero-step numerical optimization setting, the LLM is provided only with the mathematical system specifications (Hamiltonian, state targets, constraints, and time discretization) in the prompt and must propose both the functional form and the starting parameters without any subsequent SPSA parameter updates. The traces below are reproduced verbatim from the original logs, with irrelevant search steps or formatting operations replaced by $[...]$ for readability.

\subsection{Qubit Ground-State Transfer}
\label{app:trace_cd}

\noindent\textbf{System Prompt Specifications:}
\begin{small}
\begin{verbatim}
SYSTEM:
- Hamiltonian: H_g(t) = [1.0 * (1 + (t/1.2)^3)]\sigma_x 
  + [2.0 + (-2.0 - 2.0) * (10*(t/1.2)^3 - 15*(t/1.2)^4 + 6*(t/1.2)^5)]\sigma_z 
  - g(t)\sigma_y
- Goal: Transfer from ground state at t=0 to ground state at t=T
- Time Discretization: 40 slices over total time T = 1.2
\end{verbatim}
\end{small}

\noindent\textbf{Verbatim LLM Reasoning Trace:}
\begin{small}
\begin{verbatim}
[...]
Wait, the control g(t) is part of the total Hamiltonian.
If we want to follow the ground state of H_drift(t) = \Omega_x(t) \sigma^x 
+ \Omega_z(t) \sigma^z, we need to add a CD term.
The CD term for a spin in a magnetic field \vec{B}_0(t) is 
H_CD = \frac{\vec{B}_0 \times \dot{\vec{B}}_0}{2 |\vec{B}_0|^2} \cdot \vec{\sigma}.
Here \vec{B}_0 = (\Omega_x, 0, \Omega_z).
\dot{\vec{B}}_0 = (\dot{\Omega}_x, 0, \dot{\Omega}_z).
\vec{B}_0 \times \dot{\vec{B}}_0 = (0, \Omega_z \dot{\Omega}_x - \Omega_x \dot{\Omega}_z, 0).
So the CD term is along \sigma^y.
H_CD = \frac{\Omega_z \dot{\Omega}_x - \Omega_x \dot{\Omega}_z}{2 (\Omega_x^2 + \Omega_z^2)} \sigma^y.
Our control term is -g(t) \sigma^y.
So we should set -g(t) = \frac{\Omega_z \dot{\Omega}_x - \Omega_x \dot{\Omega}_z}{2 (\Omega_x^2 + \Omega_z^2)}.
Or g(t) = \frac{\Omega_x \dot{\Omega}_z - \Omega_z \dot{\Omega}_x}{2 (\Omega_x^2 + \Omega_z^2)}.
[...]
Let \theta(t) be the angle of the field in the X-Z plane.
The instantaneous eigenstates rotate in the X-Z plane. To track the ground state, 
we need a rotation around Y.
The counter-diabatic driving term required to suppress transitions is roughly 
\dot{\theta}/2 \sigma_y.
\tan \theta = \Omega_x / \Omega_z.
\dot{\theta} = \frac{\Omega_z \dot{\Omega}_x - \Omega_x \dot{\Omega}_z}{\Omega_x^2 + \Omega_z^2}.
So H_CD = \frac{\dot{\theta}}{2} \sigma_y.
Control term in H is -g(t) \sigma_y.
So -g = \dot{\theta}/2 implies g = -\dot{\theta}/2.
g(t) = \frac{\Omega_x \dot{\Omega}_z - \Omega_z \dot{\Omega}_x}{2(\Omega_x^2 + \Omega_z^2)}.
[...]
So the physics is correct.
The error must be discretization.
Midpoint rule is 2nd order. Error ~ dt^2 ~ 10^{-3}.
The CD term cancels transitions in continuous time.
In discrete time, the "kick" from H(t_k) is exact for that slice, 
but the sequence of rotations doesn't perfectly track.
[...]
\end{verbatim}
\end{small}
\subsection{NMR Coherence Transfer}
\label{app:trace_twospin}

\noindent\textbf{System Prompt Specifications:}
\begin{small}
\begin{verbatim}
SYSTEM:
- Hamiltonian: H(t) = 2*pi*J*Iz@Sz + u_Ix(t)*Ix + u_Iy(t)*Iy + u_Sx(t)*Sx + u_Sy(t)*Sy
- Coupling constant: J = 100.0 Hz
- Initial state: rho(0) = Sx
- Target state: rho(T) = Ix
- Time Discretization: 60 slices over total time T = 15.0 ms
\end{verbatim}
\end{small}

\noindent\textbf{Verbatim LLM Reasoning Trace:}
\begin{small}
\begin{verbatim}
[...]
This is a standard INEPT (Insensitive Nuclei Enhanced by Polarization Transfer) 
type transfer or simply coherence transfer via scalar coupling.
The coupling term 2\pi J I_z S_z generates anti-phase coherence.
Evolution under 2\pi J I_z S_z for time \tau:
S_x \xrightarrow{2\pi J I_z S_z \tau} S_x \cos(\pi J \tau) + 2 S_y I_z \sin(\pi J \tau).
[...]
At t = 1/(2J), \pi J t = \pi/2, \cos=0, \sin=1. State: 2 S_y I_z.
[...]
Change 90_x^I to 90_{-x}^I: 2 S_z I_z \to 2 S_z I_y.
2 S_z I_y \xrightarrow{J, 1/2J} I_x.
So:
1. Delay 5ms.
2. S: 90_x, I: 90_{-x}. (Simultaneous hard pulses)
3. Delay 5ms.
Total time 10ms + pulse widths.
[...]
The sequence R_x(90) - J(1/2J) - R_y(90) - J(1/2J) - R_x(90) is a standard 
decomposition for SWAP using Ising coupling.
[...]
Slice width \Delta t = 15/60 = 0.25 ms.
Midpoints: 0.125, 0.375, ..., 14.875.
If I define a pulse from 0.0 to 0.5 ms.
Midpoints covered: 0.125, 0.375. (2 slices).
Total duration sampled: 2 \times 0.25 = 0.5 ms.
So if I set amplitude A, effective rotation is A \times 0.5 ms.
So A \times 0.0005 = \pi/2 \implies A = 3141.6.
[...]
During the pulse, J-coupling is also active.
u \approx 6283 rad/s. 2\pi J \approx 628 rad/s. Ratio 10:1.
Error might be significant?
0.25 ms \times 628 rad/s \approx 0.157 rad \approx 9^\circ.
This is not negligible.
[...]
\end{verbatim}
\end{small}
\end{widetext}

\end{document}